# Physical Layer Security for Integrated Sensing and Communication: A Survey

Toshiki Matsumine, *Member, IEEE*, Hideki Ochiai, *Fellow, IEEE*, and Junji Shikata, *Member, IEEE*

*Abstract*—Integrated sensing and communication (ISAC) has become a crucial technology in the development of next-generation wireless communication systems. The integration of communication and sensing functionalities on a unified spectrum and infrastructure is expected to enable a variety of emerging use cases. The introduction of ISAC has led to various new challenges and opportunities related to the security of wireless communications, resulting in significant research focused on ISAC system design in relation to physical layer security (PLS). The shared use of spectrum presents a risk where confidential messages embedded in probing ISAC signals may be exposed to potentially malicious sensing targets. This situation creates a tradeoff between sensing performance and security performance. The sensing functionality of ISAC offers a unique opportunity for PLS by utilizing sensing information regarding potential eavesdroppers to design secure PLS schemes. This study examines PLS methodologies to tackle the specified security challenge associated with ISAC. The study begins with a brief overview of performance metrics related to PLS and sensing, as well as the optimization techniques commonly utilized in the existing literature. A thorough examination of existing literature on PLS for ISAC is subsequently presented, with the objective of emphasizing the current state of research. The study concludes by outlining potential avenues for future research pertaining to secure ISAC systems.

### LIST OF ABBREVIATIONS

| | |
|---|---|
| 3D | Three-Dimensional |
| A3C | Asynchronous Advantage Actor-Critic |
| ADMM | Alternating Direction Method of Multipliers |
| ADPAR | Angular-Domain Peak-to-Average Ratio |
| AFDM | Affine Frequency Division Multiplexing |
| AI | Artificial Intelligence |
| ALM | Augmented Lagrangian Method |
| AM | Arithmetic Mean |
| AN | Artificial Noise |

This work was in part supported by JSPS KAKENHI Grant Number JP22K19773. This research was in part conducted under a contract of "Research and development on new generation cryptography for secure wireless communication services" among "Research and Development for Expansion of Radio Wave Resources (JPJ000254)", which was supported by the Ministry of Internal Affairs and Communications, Japan.

T. Matsumine is with the Institute of Advanced Sciences, Yokohama National University, Yokohama, Japan (e-mail: matsumine-toshiki-mh@ynu.ac.jp).

H. Ochiai is with the Graduate School of Engineering, Osaka University, Osaka, Japan (e-mail: ochiai@comm.eng.osaka-u.ac.jp).

J. Shikata is with Graduate School of Environment and Information Sciences, Yokohama National University, Yokohama, Japan (email: shikata-junji-rb@ynu.ac.jp).

| | |
|---|---|
| AO | Alternating Optimization |
| AoA | Angle-of-Arrival |
| AoD | Angle-of-Departure |
| AP | Access Point |
| AWGN | Additive White Gaussian Noise |
| BCCD | Block Cyclic Coordinate Descent |
| BCD | Block Coordinate Descent |
| BCRB | Bayesian Cramér-Rao Bound |
| BER | Bit Error Rate |
| BF | Beamforming |
| BPG | Beampattern Gain |
| BPM | Beampattern Matching |
| BS | Base Station |
| BTI | Bernstein-Type Inequality |
| CC | Cross-Correlation |
| CCCP | Concave-Convex Procedure |
| CCP | Convex-Concave Procedure |
| CD | Coordinate Descent |
| CF | Cell-Free |
| CI | Constructive Interference |
| CoMP | Coordinated Multi-Point Transmission |
| CR | Covert Rate |
| CRB | Cramér-Rao Bound |
| CSI | Channel State Information |
| DAM | Delay Alignment Modulation |
| DC | Difference−of−Convex |
| DDPG | Deep Deterministic Policy Gradient |
| DFRC | Dual-Functional Radar-Communications |
| DI | Destructive Interference |
| DL | Downlink |
| DM | Distance-Majorization |
| DMA | Dynamic Metasurface Antenna |
| DoF | Degree of Freedom |
| DQN | Deep Q-Network |
| DRL | Deep Reinforcement Learning |
| EKF | Extended Kalman Filter |
| EM | Electromagnetic |
| EMMSE | Expected Minimum Mean Squared Error |
| ES | Energy-Splitting |
| FAS | Fluid Antenna System |
| FC | Fully-Connected |
| FD | Full-Duplex |
| FH | Frequency Hopping |
| FIM | Flexible Intelligent Metasurface |
| FOTE | First-Order Taylor Expansion |
| FP | Fractional Programming |
| FPA | Fixed-Position Antenna |
| FPJ | Fully-Passive Jamming |
| GBD | Generalized Benders Decomposition |
| GLRT | Generalized Likelihood Ratio Test |
| GM | Geometric Mean |
| HAPS | High-Altitude Platform Station |
| HD | Half-Duplex |
| HFCS | Hopping Frequency Combination Selection |
| HFPS | Hopping Frequency Permutation Selection |
| HMIMO | Holographic MIMO |
| HPA | High Power Amplifier |
| IM | Index Modulation |
| IOS | Intelligent Omni-Surface |
| IoT | Internet of Things |
| IoV | Internet of Vehicles |
| ISAC | Integrated Sensing and Communication |
| ISI | Inter-Symbol Interference |
| ISLR | Integrated Sidelobe Level Ratio |
| ISMR | Integrated Sidelobe-to-Mainlobe Ratio |
| KPI | Key Performance Indicator |
| LD | Lagrange Duality |
| LDPC | Low-Density Parity Check |
| LEO | Low-Earth Orbit |
| LMI | Linear Matrix Inequality |
| LOS | Line-of-Sight |
| LS | Least Squares |
| LSTM | Long Short-Term Memory |
| MA | Movable Antenna |
| MAC | Multiple-Access Channel |
| MC | Multi-Carrier |
| MI | Mutual Information |
| MILP | Mixed-Integer Linear Programming |
| MIMO | Multi-Input Multi-Output |
| MLP | Multi-Layer Perceptron |
| MM | Majorization-Minimization |
| MMSE | Minimum Mean Squared Error |
| mmWave | Millimeter Wave |
| MO | Manifold Optimization |
| MS | Mode Switching |
| MSE | Mean Squared Error |
| MUI | Multi-User Interference |
| NLOS | Non-line-of-Sight |
| NOMA | Non-Orthogonal Multiple Access |
| OFDM | Orthogonal Frequency Division Multiplexing |
| OTFS | Orthogonal Time-Frequency Space |
| PAPR | Peak-to-Average-Power Ratio |
| PCRB | Posterior Cramér-Rao Bound |
| PD | Power Domain |
| PGD | Projected Gradient Descent |
| PLS | Physical Layer Security |
| PPO | Proximal Policy Optimization |
| PSK | Phase-Shift Keying |
| QAM | Quadrature Amplitude Modulation |
| QoS | Quality Of Service |
| RCCE | Radar and Communication Coexisting |
| RCG | Riemannian Conjugate Gradient |
| RCS | Radar Cross Section |
| RF | Radio Frequency |
| RHS | Reconfigurable Holographic Surface |
| RIS | Reconfigurable Intelligent Surface |
| RL | Reinforcement Learning |
| RM | Rank Minimization |
| RSMA | Rate-Splitting Multiple Access |
| SAC | Soft Actor-Critic |
| SCA | Successive Convex Approximation |
| SCNR | Signal-to-Clutter-plus-Noise Ratio |
| SCP | Sequential Convex Programming |
| SDP | Semidefinite Programming |
| SDR | Semidefinite Relaxation |
| SER | Symbol Error Rate |
| SIC | Successive Interference Cancellation |
| SIM | Stacked Intelligent Metasurfaces |
| SINR | Signal-to-Interference-plus-Noise Ratio |
| SNR | Signal-to-Noise Ratio |
| SOA | Sequential Optimization Algorithm |
| SOC | Second-Order Cone |
| SOP | Secrecy Outage Probability |
| SOTE | Second-Order Taylor Expansion |
| SPSA | Simultaneous Perturbation Stochastic Approximation |
| SR | Secrecy Rate |
| SROCR | Sequential Rank-One Constraint Relaxation |
| SSR | Semantic Secrecy Rate |
| STAR | Simultaneously Transmitting and Reflecting |
| SWIPT | Simultaneous Wireless Information And Power Transfer |
| TDD | Time Division Duplex |
| TDP | Target Detection Probability |
| THz | Terahertz |
| TIP | Target Illumination Power |
| TRPO | Trust Region Policy Optimization |
| TS | Time Switching |
| UAV | Unmanned Aerial Vehicle |
| UL | Uplink |
| ULA | Uniform Linear Array |
| V2I | Vehicle-To-Infrastructure |
| V2V | Vehicle-To-Vehicle |
| V2X | Vehicle-To-Everything |
| XL | Extremely Large-Scale |
| ZF | Zero-Forcing |

# I. INTRODUCTION

Integrated sensing and communication (ISAC) has received increasing attention in recent years as a result of its combination of sensing and communication within both hardware architecture and signal processing algorithms [1]–[10]. ISAC has been recognized as a crucial facilitator for various use cases within IMT-2030, which is anticipated to be a significant milestone in the development of 6G [11]. The sensing functionality of ISAC is expected to enable advanced applications, including high precision positioning and localization of devices and objects [12], real-time 3D-mapping for automated transport [13], and the development of digital twins [14]. As a result, the development of ISAC technologies is anticipated to play a crucial role in facilitating a significant transition to innovative environment-aware communications [15], [16], which utilize prior knowledge of the existing local environment for the design and optimization of communication networks.

Wireless communication and sensing systems are inherently exposed to potential security threats due to the broadcast nature of wireless channels. ISAC systems present more complex security challenges than communication-only systems. These challenges can be classified into two categories: communication security and sensing security [17]. The former arises from the possibility that the embedded information within the ISAC waveform may be exposed to sensing targets, which could function as eavesdroppers. The inherent property of ISAC establishes a fundamental tradeoff between information confidentiality and sensing performance in target detection [18]. In contrast, passive sensing eavesdroppers could exploit the transmitted ISAC waveform reflected by targets to extract sensing information regarding those targets or their surrounding environment, posing significant challenges in ensuring the security of the sensing functionality within ISAC systems.

Physical layer security (PLS) techniques are a potential solution to the aforementioned communication security issue of ISAC systems [19]–[30]. Unlike a cryptographic technique, PLS provides information-theoretic security by utilizing the unpredictability of wireless channels, making it immune to attacks from both classical and quantum computers. Nevertheless, existing PLS techniques often require ideal assumptions about the eavesdropper's channel information. For instance, it is frequently assumed that the channel state information (CSI) of the eavesdropper's channel is accessible at the transmitter side. However, this is only applicable when the eavesdropper is active or when the eavesdropper is a licensed user. In practical scenarios where the eavesdropper is passive, the transmitter will not be able to acquire the requisite CSI to accurately characterize the eavesdropper's channel. This limitation may hinder the effective implementation of PLS in practical systems.

The sensing functionality of ISAC presents a viable approach to addressing the challenges associated with eavesdropping during the transmission of secret messages. Specifically, the sensing capabilities of ISAC can be utilized to estimate the directions of potential eavesdroppers [31], and the gathered information can be employed to optimize transmission strategies, thereby enhancing the security level. Given the unique security challenges and opportunities presented by ISAC, PLS techniques for ISAC systems have been the focus of extensive research in the literature, evidenced by a marked rise in related publications in recent years. Consequently, it is imperative to systematically categorize and summarize these publications to illuminate the current state and technical challenges.

### A. Related Works

A substantial number of survey, tutorial, and magazine papers on general ISAC have been published recently [1]–[8], [10], [18], [32]–[41]. Nevertheless, despite its importance, there exists a limited number of survey papers focused on PLS problems within the context of ISAC. For instance, recent publications [18], [42]–[45] have addressed the security concerns related to ISAC. In [18], the authors presented a succinct overview of the unique security challenges and opportunities associated with ISAC, along with a discussion of the challenges related to PLS design for ISAC systems.

More recent studies have examined the security and privacy concerns associated with ISAC [42], [44]. In [42], the authors examined practical solutions addressing security and privacy concerns in ISAC, specifically focusing on defensive strategies to counter adversarial sensing and location-sensitive spoofing attacks. In [44], the authors discussed the technical challenges associated with facilitating privacy-preserving and secure ISAC networks, and presented a framework designed for security and privacy preservation within the network. This framework strengthens network defenses and improves ISAC operational capabilities through the utilization of artificial intelligence (AI), friendly jamming, and reconfigurable intelligent surfaces (RISs). In addition, the authors of [43] explored security vulnerabilities in ISAC systems at several levels, including hardware, omniscient, and application layers. While the survey explored PLS techniques for ISAC as a hardware layer security, including the usage of RIS and artificial noise (AN) to maximize a secrecy rate (SR), it was very brief on PLS issues.

More recently, the authors of [45] provided an overview of PLS techniques for ISAC. Specifically, they discussed several PLS techniques, such as AN and symbol-wise precoding to ensure communication security, and beamforming (BF) techniques to address sensing privacy issues. They also introduced sensing-assisted PLS and covert communication schemes that exploit the sensing functionality of ISAC to design secure transmission strategies, as well as several open challenges. Furthermore, the authors of [46] provided a survey of PLS techniques for ISAC, focusing on RIS-based approaches.

### B. Our Contributions

In contrast to the aforementioned works, we aim to present a comprehensive survey of existing articles on PLS in ISAC systems. To the best of our knowledge, this is the first survey work that covers a wide range of PLS approaches for ISAC systems. Based on a detailed examination of recent improvements in PLS technologies for ISAC, we propose various future research directions in the realm of secure ISAC systems. In summary, we made the following contributions:

- A series of key performance metrics for sensing and PLS is presented, accompanied by various optimization techniques that have been widely implemented to tackle PLS challenges in ISAC systems.
- A comprehensive survey of the state-of-the-art PLS techniques for ISAC is conducted, highlighting their interaction with other key technologies, including RIS, non-orthogonal multiple access (NOMA), and unmanned aerial vehicle (UAV).
- Several prospective research directions in secure ISAC systems are presented in order to encourage comprehensive investigation of this subject.

The organization of this paper is illustrated in Fig. 1. In Section II, we introduce performance metrics for both communication and sensing, as well as optimization techniques that are frequently adopted in the literature. Section III presents a detailed survey of PLS techniques for ISAC, categorizing them according to the specific scenarios considered. In Section IV, we summarize the technical challenges and future research directions. Finally, concluding remarks are given in Section V.

*Notation:* Matrices, vectors, and scalars are denoted by bold uppercase letters (e.g., $\mathbf{X}$), bold lowercase letters (e.g., $\mathbf{x}$), and normal font (e.g., $x$), respectively. The symbols $(\cdot)^T$ and $(\cdot)^H$ denote the transpose and the Hermitian transpose of the matrix, respectively, $|\cdot|$ and $\|\cdot\|$ denote the absolute value and the Euclidean norm, respectively, and $\mathbb{E}\{\cdot\}$ denotes the statistical expectation. Also, $\mathbf{I}_N$ is the identity matrix of size $N$ and $[x]^+ \triangleq \max(x, 0)$.

## II. PERFORMANCE METRICS AND OPTIMIZATION TECHNIQUES

In this section, we first describe three primary design strategies of ISAC and then introduce several key performance measures for PLS and sensing, illustrated with a system example. Subsequently, we present several techniques to address the system optimization problems.

### A. Design Principles

The design of waveforms that satisfy the requirements of both communication and sensing functionalities plays a vital role in ISAC systems. The design of ISAC waveforms is typically executed through three methodologies: sensing-centric design, communication-centric design, and joint design approaches [5], [37].

*1) Communication-Centric Design:* The communication-centric design is predicated on the exploitation of existing communication waveforms and systems for the implementation of sensing functionality. The approach prioritizes communication while extracting target information from radar echoes with minimal modifications.

A notable example is the orthogonal frequency division multiplexing (OFDM) waveform [47], due to its compatibility with the 4G and 5G standards. In addition, orthogonal time-frequency space (OTFS) modulation has been considered as a potential candidate waveform for ISAC applications due to the inherent connection of the delay-Doppler parameters to sensing applications that enables direct conversion of delay and Doppler shifts into ranges and speeds of sensing targets [48]–[50]. More recently, the application of affine frequency division multiplexing (AFDM) modulation [51], [52] was considered in ISAC [53]–[55]. While in principle, any communication waveform can be leveraged for ISAC, this design may lead to degraded sensing performance, as these waveforms are not designed for sensing.

*2) Sensing-Centric Design:* Sensing-centric design is an approach that aims to incorporate communication symbols into radar waveforms, also known as the information embedding method [56], [57]. For example, this can be achieved by modulating communication symbols into the amplitude, phase, and frequency of the chirp waveform [58]–[60]. Another significant approach involves information embedding in spatial domain for a multi-input multi-output (MIMO) radar system, including sidelobe control [56] and array modulation [61]–[63]. Furthermore, index modulation (IM) has emerged as a promising technique, whereby communication symbols are modulated into different indices of antennas and frequencies [64], [65]. However, the maximum bit rate of most sensing-centric approaches is constrained by the pulse repetition frequency (PRF), resulting in low data rates [5], [66].

*3) Joint Design:* The joint design aims to achieve enhanced trade-off between sensing and communication performance by solving a multi-objective optimization, where both sensing and communication performance metrics are used as objective functions or constraints. In general, the joint design approach provides significant flexibility and high degree of freedom (DoF), potentially offering a better tradeoff between the two functionalities. On the other hand, the optimization problem results in higher complexity compared to the other two designs.

The primary focus of this paper is on the joint design of ISAC, which aims to achieve a scalable performance trade-off between PLS and sensing. To this end, performance metrics for both PLS and sensing are introduced based on a system example in the following.

### B. An Example of Generic ISAC System

Let us consider a MIMO-dual-functional radar-communications (DFRC) system with mono-static sensing[1], as illustrated in Fig. 2. The system under consideration involves a base station (BS) is equipped with a uniform linear array (ULA) containing $N_t$ transmit and $N_r$ receive antennas[2]. The BS simultaneously serves $K$ single-antenna communication users and detects $T$ point-like targets at the specified location of interest. These targets are classified as potential eavesdroppers, with the primary aim of intercepting information transmissions occurring between the base station and legitimate users. This scenario has also been examined in the existing literature, e.g., [31], [70], [71].

*1) Communication Signal Model:* Let $\mathbf{X} \in \mathbb{C}^{N_t \times L}$ denote the transmit waveforms, where $L$ is the number of time-domain snapshots. By transmitting the dual-functional waveforms to $K$ communication users, the received signal matrix at the receivers can be expressed as

$$\mathbf{Y}_C = \mathbf{H}\mathbf{X} + \mathbf{N}_C, \tag{1}$$

where $\mathbf{H} = [\mathbf{h}_1, \ldots, \mathbf{h}_K]^H \in \mathbb{C}^{K \times N_t}$ represents the communication channel matrix, and $\mathbf{N}_C \in \mathbb{C}^{K \times L}$ is the additive white Gaussian noise (AWGN) matrix with the variance of each entry being $\sigma_c^2$.

We consider a data matrix given as

$$\mathbf{S} = \begin{bmatrix} \mathbf{S}_C \\ \mathbf{S}_A \end{bmatrix} \in \mathbb{C}^{(K+N_t) \times L}, \tag{2}$$

where $\mathbf{S}_C \in \mathbb{C}^{K \times L}$ represents the unit-power data stream intended to communication users and $\mathbf{S}_A \in \mathbb{C}^{N_t \times L}$ signifies the dedicated probing streams, which is orthogonal to $\mathbf{S}_C$, i.e., $\mathbb{E}\{\mathbf{S}\mathbf{S}^H\} = \mathbf{I}_{K+N_t}$. Also, we define the BF matrix as

$$\mathbf{W} = [\mathbf{w}_1, \ldots, \mathbf{w}_{K+N_t}] = [\mathbf{W}_C, \mathbf{W}_A] \in \mathbb{C}^{N_t \times (K+N_t)}, \quad (3)$$

where $\mathbf{W}_C \in \mathbb{C}^{N_t \times K}$ and $\mathbf{W}_A \in \mathbb{C}^{N_t \times N_t}$ are the communication and radar BF matrices, respectively. Then, the transmit waveform $\mathbf{X}$ in (1) is expressed as[3]

$$\mathbf{X} = \mathbf{W}\mathbf{S} = \mathbf{W}_C\mathbf{S}_C + \mathbf{W}_A\mathbf{S}_A. \quad (4)$$

---

[1]For mono-static sensing with a co-located transmitter and receiver pair, self-interference becomes an practical issue due to full-duplex (FD) operation [5]. In this paper, we assume that the self-interference is ideally suppressesd by cancellation techniques [67], [68].

[2]On the other hand, sensing with transmitter and receiver placed at different locations is referred to as bi-static sensing. More generally, a radar configuration involving multiple transmitters and receivers distributed over different locations is referred to as multi-static sensing [69].

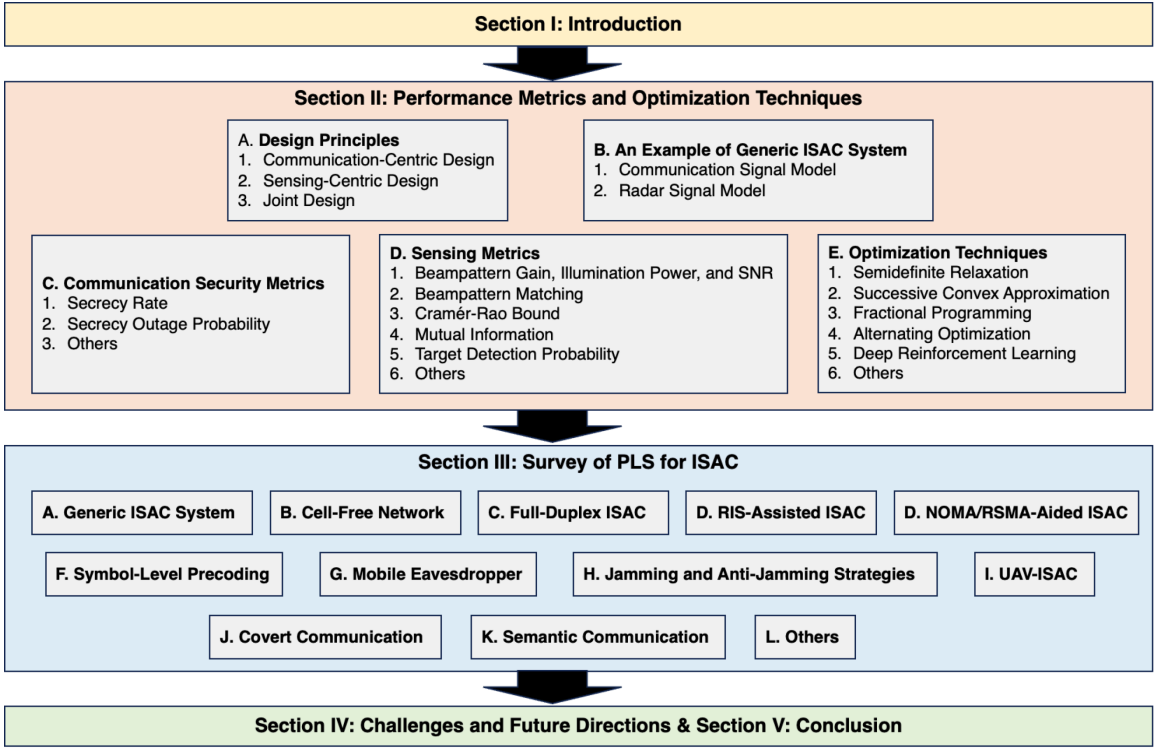

Fig. 1. The organization of this paper.

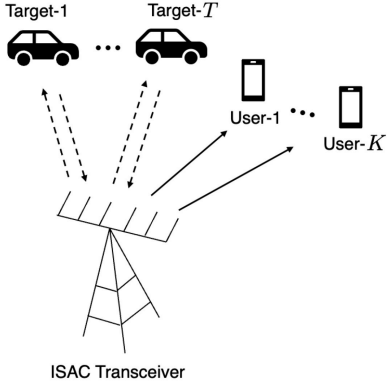

Fig. 2. A generic ISAC system with $K$ communication users and $T$ sensing targets.

[3] The inclusion of the dedicated sensing waveform $\mathbf{W}_A\mathbf{S}_A$ provides the extra DoF in the beampattern design and improves the sensing performance [5]. This term is often replaced with an AN term in the context of PLS.

Furthermore, the signal-to-interference-plus-noise ratio (SINR) of the $i$-th user is given by

$$\text{SINR}_i^U = \frac{|\mathbf{h}_i^H \mathbf{w}_i|^2}{\sum_{\substack{k=1 \\ k \neq i}}^{K} |\mathbf{h}_i^H \mathbf{w}_k|^2 + \|\mathbf{h}_i^H \mathbf{W}_A\|^2 + \sigma_C^2}. \quad (5)$$

*2) Radar Signal Model:* Let $\{\theta\}_{t=1}^T$ denote the angle parameters of targets of interest. By emitting the waveform $\mathbf{X}$ to sense eavesdroppers, the reflected echo signal matrix at the BS receive array is given as

$$\mathbf{Y}_R = \underbrace{\sum_{t=1}^{T} \mathbf{a}(\theta_t)\beta_t \mathbf{b}^H(\theta_t)}_{\triangleq \mathbf{G}} \mathbf{X} + \mathbf{N}_R, \quad (6)$$

where $\mathbf{G} \in \mathbb{C}^{N_r \times N_t}$ is the target response matrix, $\mathbf{a}(\theta_t) \in \mathbb{C}^{N_r \times 1}$ and $\mathbf{b}(\theta_t) \in \mathbb{C}^{N_t \times 1}$ are the steering vectors for the receive and transmit arrays, respectively, and $\beta_t$ is the complex amplitude proportional to the radar cross section (RCS) of the $t$-th eavesdropper. Given the azimuth angle of the target relative to the BS, $\theta$, the receive steering vector is defined as

$$\mathbf{a}(\theta) = \left[1, e^{-j2\pi\Delta\sin\theta}, \ldots, e^{j2\pi(N_r-1)\Delta\sin\theta}\right]^T, \quad (7)$$

where $\Delta \triangleq d/\lambda$, $d$ and $\lambda$ are the interval between adjacent antennas and the wavelength, respectively. The transmit steering vector $\mathbf{b}(\theta)$ is also defined in the same way. Finally, $\mathbf{N}_R \in \mathbb{C}^{N_r \times L}$ is an AWGN matrix with the variance of each entry being $\sigma_R^2$.

The eavesdropping SINR received at the $t$-th eavesdropper with respect to the $i$-th communication user is written as

$$\text{SINR}_{t,i}^{\text{E}} = \frac{|\alpha_t \mathbf{b}^H(\theta_t)\mathbf{w}_i|^2}{\sum_{\substack{k=1 \\ k \neq i}}^{K} |\alpha_t \mathbf{b}^H(\theta_t)\mathbf{w}_k|^2 + \|\alpha_t \mathbf{b}^H(\theta_t)\mathbf{W}_A\| + \sigma_E^2} \tag{8}$$

where $\alpha_t$ denotes the complex path-loss coefficient of the $t$-th target and $\sigma_E^2$ denotes the variance of AWGN received by each eavesdropper.

A significant challenge in the design of ISAC systems involves optimizing the transmit BF $\mathbf{W}$, which must effectively perform dual functions: target sensing and communication. In what follows, we present essential performance metrics related to PLS and sensing.

### C. Communication Security Metrics

In the context of PLS, several performance metrics exist to measure security performance. We present the primary performance metrics that are frequently referenced in existing literature.

*1) Secrecy Rate:* The achievable SR for the legitimate user is defined as the difference between the rates that can be achieved at the legitimate receivers and those at the eavesdroppers. The *worst-case* SR is expressed as [31]

$$R_S = \min_{i,t}[R_i^{\text{U}} - R_{t,i}^{\text{E}}]^+, \tag{9}$$

where $R_i^{\text{U}}$ and $R_{t,i}^{\text{E}}$ represent the achievable transmission rate of the $i$-th user and that of the $t$-th eavesdropper, respectively, and they are expressed as

$$R_i^{\text{U}} = \log(1 + \text{SINR}_i^{\text{U}}), \ \ R_{t,i}^{\text{E}} = \log(1 + \text{SINR}_{t,i}^{\text{E}}). \tag{10}$$

*2) Secrecy Outage Probability:* The secrecy outage probability (SOP) serves as a relevant metric for quantifying the likelihood that secure transmission is unattainable [72], [73]. The SOP is defined as the likelihood of a secrecy outage event occurring. A commonly accepted definition of secrecy outage refers to the occurrence in which the instantaneous secrecy capacity, denoted as $R_S$, falls below a specified target SR, represented as $\Gamma_S$ [74]–[76], i.e.,

$$P_{\text{out}}(\Gamma_S) \triangleq \Pr(R_S < \Gamma_S). \tag{11}$$

The secrecy outage is thus defined as a situation in which a message fails to be reliably received by the intended recipient or when information is compromised and accessed by an unauthorized eavesdropper.

*3) Others:* In addition to the SR and SOP, the SINR of the communication user and the target in (5) and (8) is frequently employed as an optimization objective or constraint. Furthermore, the bit error rate (BER) or the symbol error rate (SER) of the user and the eavesdropper is employed to evaluate the reliability of communication and the potential for information leakage.

In the realm of covert communication [77], a covert rate (CR) serves as a key performance metric. A CR is characterized as a feasible rate that can be attained while adhering to a covert constraint, specifically regarding the minimum detection error probability of the warden.

### D. Sensing Metrics

Several key metrics for sensing performance are introduced below.

*1) Beampattern Gain, Illumination Power, and SNR:* The primary objective of waveform and BF design in radar systems is to optimize total spatial power at designated target locations [78]. This method entails utilizing the available transmit power to optimize the probing signal power (beampattern gain (BPG)) at the target sites while minimizing it in other areas.

More specifically, given the angle of the $t$-th target $\theta_t$, the transmit BPG is defined as

$$P(\theta_t, \mathbf{R}_X) = \mathbb{E}\{\|\mathbf{b}^H(\theta_t)\mathbf{X}\|^2\} = \mathbf{b}^H(\theta_t)\mathbf{R}_X \mathbf{b}(\theta_t), \tag{12}$$

where $\mathbf{R}_X$ is the covariance matrix of the transmitted signal and can be written as

$$\mathbf{R}_X = \mathbb{E}\{\mathbf{X}\mathbf{X}^H\} = \mathbf{W}_C\mathbf{W}_C^H + \mathbf{W}_A\mathbf{W}_A^H \tag{13}$$

Then we may be interested in maximizing the cumulated power of the probing signals at the target location, which is given by [78]

$$\sum_{t=1}^{T} P(\theta_t, \mathbf{R}_X). \tag{14}$$

A radar performance metric that is analogous in nature is target illumination power (TIP). This is essentially equivalent to the transmit BPG in (12) and is defined as [79]

$$P_{\text{ill}}(\theta_t, \mathbf{R}_X) = \mathbb{E}\{\|\alpha_t \mathbf{b}^H(\theta_t)\mathbf{X}\|^2\}. \tag{15}$$

Furthermore, the signal-to-noise ratio (SNR) or SINR at the radar receiver has been widely utilized as a sensing metric in the existing literature. This metric is affected by ambient conditions, such as interference and noise. Let $\mathbf{u} \in \mathbb{C}^{N_t}$ denote a radar receive filter [80] to suppress the interference from the BS and noise. The filtered signal is given by

$$\mathbf{u}^H\mathbf{Y}_R = \mathbf{u}^H\mathbf{G}(\mathbf{W}_C\mathbf{S}_C + \mathbf{W}_A\mathbf{S}_A) + \mathbf{u}^H\mathbf{N}_R, \tag{16}$$

and the radar output SINR can be expressed as

$$\text{SINR}^R = \frac{\|\mathbf{u}^H\mathbf{G}\mathbf{W}_A\|^2}{\|\mathbf{u}^H\mathbf{G}\mathbf{W}_C\|^2 + \mathbf{u}^H\sigma_R^2\mathbf{I}_{N_t}\mathbf{u}}. \tag{17}$$

In most practical applications, sensing requires an unobstructed line-of-sight (LOS) pathway between the sensor and the observed object. In radar applications, signals that are reflected by objects not corresponding to the targets of interest are classified as clutter. This clutter is deemed harmful and requires mitigation. In these scenarios, the signal-to-clutter-plus-noise ratio (SCNR) serves as a metric for evaluating the performance of sensing systems, aiming to maximize this ratio [5], [81].

*2) Beampattern Matching:* While the waveform design based on BPG in (14) maximizes the total power at the locations of the targets of interest, the resulting powers at the target locations can vary significantly. In contrast, the beampattern matching (BPM) [71], [78], [82], [83] accommodates the uniform elemental transmit power constraint and allows for the control of the power at each target location.

The objective of BPM is to approximate or match the desired transmit beampattern over the specified sectors in a least squares (LS) sense. To this end, the mean squared error (MSE) between the designed and desired beampatterns is formulated as

$$L_{\text{MSE}}(\mathbf{R}_X, \gamma) = \frac{1}{\ell} \sum_{l=1}^{\ell} |\gamma \Phi(\mu_l) - P(\mu_l; \mathbf{R}_X)|^2, \qquad (18)$$

where $\gamma$ is a scaling factor to be optimized, $\Phi(\mu_l)$ is the desired beampattern, and $\{\mu_l\}_{l=1}^{\ell}$ is a fine grid of points that cover the location sectors of interest. Let $\Delta$ denote the width of desired beampattern; then the desired beampattern is given by [83]–[85]

$$\Phi(\theta) = \begin{cases} 1 & |\theta - \theta_t| \leq \frac{\Delta}{2}, \exists t \in \{1, \dots, T\}, \\ 0 & \text{otherwise.} \end{cases} \qquad (19)$$

It is also possible that this criterion may include an additional term that penalizes large values of the cross-correlation (CC) beampattern [71], [78], [83], resulting in a weighted optimization of the MSE between the desired and design beampattern and the CC. Specifically, the mean squared CC pattern is given as

$$L_{\text{CC}}(\mathbf{R}_X) = \frac{T}{T^2 - T} \sum_{t=1}^{T-1} \sum_{q=t+1}^{T} |P_{\text{CC}}(\theta_t, \theta_q; \mathbf{R}_X)|^2. \qquad (20)$$

Also, the CC pattern between direction $\theta_1$ and $\theta_2$ is defined as

$$P_{\text{CC}}(\theta_1, \theta_2; \mathbf{R}_X) = \mathbf{b}^H(\theta_2) \mathbf{R}_X \mathbf{b}(\theta_1). \qquad (21)$$

The summation is normalized by $\frac{T}{T^2-T}$ since there exist $\frac{T^2-T}{2}$ pairs of distinct directions in the set $\{\theta_t\}_{t=1}^{T}$.

From (18) and (20), the total loss function for radar waveform design is given as

$$L_{\text{R}}(\mathbf{R}_X, \gamma) = L_{\text{MSE}}(\mathbf{R}_X, \gamma) + w L_{\text{CC}}(\mathbf{R}_X), \qquad (22)$$

where $w$ is a weighting factor to balance the two objectives.

*3) Cramér-Rao Bound:* The MSE is a possible performance metric for measuring the accuracy of an estimator. The MSE is defined as the mean of the squared error between the true value of a parameter $\theta$ and its estimate $\hat{\theta}$. The Cramér-Rao bound (CRB) is a foundational concept in estimation theory that establishes a lower bound on the variance of any unbiased estimator of a parameter $\theta$. The CRB is defined as the inverse of the Fisher information matrix [7], [86]–[89]. The Fisher information matrix is defined as the expectation of the curvature (negative second derivative) of the likelihood function with respect to $\theta$. This curvature is a measure of the "sharpness" or the accuracy of the estimator.

Furthermore, the target location distribution may be known *a priori* based on empirical data or target movement patterns [90]. By properly exploiting such prior information, effective sensing can be completed in a single iteration instead of iteratively over multiple time slots. A metric to characterize the sensing performance that exploits such prior distribution information is the posterior Cramér-Rao bound (PCRB) or Bayesian Cramér-Rao bound (BCRB) [91], which quantifies a global lower bound for the MSE of unbiased estimators. In light of this, the study [92] established the PCRB when estimating the location parameter of a target using a MIMO radar system that utilizes prior distribution information.

*4) Mutual Information:* Mutual information (MI) is an information-theoretic metric that has garnered significant attention in the field of radar system design [93]–[97]. The sensing MI is defined as the extent to which information can be extracted from the environment from an information-theoretic perspective.

Based on (6), the radar MI between the reflected waveforms given the knowledge of transmitted waveforms and the target response is defined as [17], [98], [99],

$$I(\mathbf{Y}_{\text{R}}; \mathbf{G} | \mathbf{X}) = H(\mathbf{Y}_{\text{R}} | \mathbf{X}) - H(\mathbf{Y}_{\text{R}} | \mathbf{G}, \mathbf{X}), \qquad (23)$$

where $H(\mathbf{Y}_{\text{R}} | \mathbf{X})$ denotes the conditional differential entropy of $\mathbf{Y}_{\text{R}}$ given $\mathbf{X}$ and $H(\mathbf{Y}_{\text{R}} | \mathbf{G}, \mathbf{X})$ denotes the conditional differential entropy of $\mathbf{Y}_{\text{R}}$ given $\mathbf{G}$ and $\mathbf{X}$.

Several papers have proven that maximizing the conditional MI in (23) improves performance in target detection, parameter estimation, and related areas [100], [101]. Additionally, the authors of [102] showed that MI maximization corresponds to the minimum mean squared error (MMSE) in MIMO radar waveform design for a linear-Gaussian channel.

*5) Target Detection Probability:* A target detection probability (TDP) is frequently considered as a radar performance metric [103]–[105]. The detection task is to ascertain the presence of a target within the area of interest. In the context of PLS, applications for detecting radar targets include unlicensed target detection/monitoring. The target detection problem can be modeled as a binary hypothesis test, where $\mathcal{H}_0$ and $\mathcal{H}_1$ represent the hypotheses of target absent and target present, respectively. The detection can be performed using a generalized likelihood ratio test (GLRT) [5], [104]. The detection probability, which is defined as the probability that $\mathcal{H}_1$ holds true and the detector chooses $\mathcal{H}_1$. The false-alarm probability is defined as the probability that $\mathcal{H}_0$ holds true but the detector chooses $\mathcal{H}_1$.

*6) Others:* In addition to the previously mentioned metrics, the integrated sidelobe level ratio (ISLR) or integrated sidelobe-to-mainlobe ratio (ISMR) is a performance measure for sensing, defined as the ratio of sidelobe energy to mainlobe energy [106]–[109]. By minimizing this metric, the transmit energy is expected to be concentrated in the mainlobe region as much as possible, thereby leading to better radar-sensing performance.

Another important requirement for both sensing and communication is peak-to-average-power ratio (PAPR) of waveform. It is noted that high PAPR could result in serious nonlinear distortion of transmit signals incurred in a high power amplifier (HPA) of the radio frequency (RF) front end. Therefore, design of low PAPR waveforms has been studied for ISAC systems [39], [110]–[113].

Furthermore, in the context of a mobile eavesdropper, several studies have sought to predict the trajectory and CSI of the eavesdropper by employing the extended Kalman filter (EKF) technique. In such scenarios, the tracking MSE has been considered as a metric of sensing performance [114], [115].

While radar systems traditionally utilize deterministic signals for target sensing, ISAC systems must adopt random signals to convey information. Notably, the works [116] and [117] pioneered the introduction of the expected minimum mean squared error (EMMSE) and ergodic linear MMSE [117], respectively, for sensing with random signals. The ergodic linear MMSE is defined as the estimation error averaged over random signal realizations.

### E. Optimization Techniques

The joint design of ISAC systems in terms of the aforementioned PLS and sensing metrics generally results in non-convex optimization, which is challenging to solve directly. In the following sections, we will briefly introduce some techniques that have been widely employed in the literature to address non-convex optimization problems.

*1) Semidefinite Relaxation:* Semidefinite relaxation (SDR) is a technique used to approximate or "relax" a harder, typically non-convex optimization problem by converting it into a semidefinite programming (SDP) problem [118]. An SDP problem is a convex optimization problem where the objective function is linear, and the constraints involve semidefinite matrices. A second-order cone (SOC) programming is a subclass of SDP [119], where each constraint is defined over a SOC. These problems can be efficiently solved via interior point methods [120] or the convex optimization toolbox, such as CVX [121].

The advent of SDR has profoundly transformed the landscape of signal processing and communications, particularly in the domain of wireless communication systems. Numerous empirical evidences have demonstrated the efficacy of SDR in delivering precise approximations, a notable example being its application in multi-user and MIMO detection [122]–[126]. Additionally, recent studies have explored the application of SDR in various fields, including waveform/BF design in radar [5], [83], [89] and BF design in RIS-aided wireless networks [127]–[129].

*2) Successive Convex Approximation:* Successive convex approximation (SCA) is an iterative optimization technique used to solve a non-convex problem with a non-convex objective function and/or non-convex constraints [130][4]. In particular, at each iteration of, the non-convex components of the objective function or constraints are approximated with a convex surrogate function. Subsequently, the approximated problem is solved through the utilization of standard convex optimization techniques. This iterative process is repeated until convergence is achieved. A common surrogate function for a concave objective or constraint function is its first-order Taylor series expansion. The concept of SCA has been employed in the domain of wireless communication system design, e.g., RIS [129], [132], [133], and ISAC [134]–[136].

The principle of SCA is analogous to majorization-minimization (MM) [137], [138], which can be regarded as a more specific implementation of the overarching concept of SCA. SCA and MM aim to substitute the non-convex problem with a convex or more manageable surrogate problem and to solve a sequence of convex problems that approximate the original non-convex problem. A distinguishing feature of MM lies in its emphasis on constructing upper bounds (majorizers) for the objective function. In contrast, SCA can encompass a broader spectrum of convex approximations, not necessarily limited to upper bounds. Additionally, a distance-majorization (DM) algorithm [139] is a special case of a MM algorithm, wherein the majorizing function is explicitly constructed using a distance measure or divergence.

The concave-convex procedure (CCCP) method [140] and the convex-concave procedure (CCP) method [141] are special instances of SCA that specifically apply to difference−of−convex (DC) programming problems [142], where the objective function is the difference between a convex function and a concave function. The CCCP and CCP procedures function by linearizing the convex and concave parts, respectively, using its first-order Taylor expansion (FOTE) at the current solution and solving the resulting convex optimization problem. These operations are repeated until convergence.

*3) Fractional Programming:* Fractional programming (FP) is a specific class of optimization problems involving ratio terms, and it plays a vital role in the design and optimization of wireless communication systems due to the ubiquitous fractional structure of various performance metrics related to communication links [143]. Notably, the SINR, which is naturally defined by a fractional function, is an essential quantity for the performance evaluation of wireless communication systems. Additionally, energy efficiency, defined as the ratio between the amount of transmitted data and consumed energy, is a critical performance metric in the design of wireless communication systems [144], [145]. In addition, FP has been applied to power control in wireless cellular networks, BF to maximize the weighted sum rate of MIMO wireless cellular networks, and so forth [146].

The early works on FP primarily centered on single-ratio problems, particularly concave-convex single-ratio maximization problems, where the objective function contains a single ratio term with a nonnegative concave numerator and a positive convex denominator. Two classic techniques have been employed to address single-ratio FP problems: the Charnes-Copper transform [147] and the Dinkelbach's transform [148]. These techniques have been extensively applied to solve energy efficiency maximization problems for wireless communication systems [144], [149], [150]. While these techniques have proven effective for single-ratio FP, they are not readily applicable to multiple-ratio cases, which are more prevalent in system-level communication network design. This limitation is due to the fact that the overall system performance typically involves multiple ratio terms, such as SINRs. A notable recent advancement in the field of FP is the development of a quadratic transform and a Lagrange dual transform for solving multi-ratio FP problems [146], [151].

*4) Alternating Optimization:* Alternating optimization (AO) is an iterative procedure to optimize a multivariate function by breaking it down into a series of simpler sub-problems. This process involves optimizing over one block of function parameters while keeping the others fixed and then alternating this process among the parameter blocks. AO works best in two situations. First, when solving the subproblems is easier than solving the joint optimization problem itself. Second, when there is a natural partitioning of parameters. It is an experimentally effective and broadly applicable technique for solving optimization problems with coupled variables. Conversely, when there is no natural way to divide the problem into smaller subproblems, a sequential optimization algorithm (SOA) can be used. This algorithm typically updates all variables at each iteration based on some approximation or simplification of the original problem.

---

[4]Sequential convex programming (SCP) [131] is a very similar technique to SCA, and they are often used interchangeably.

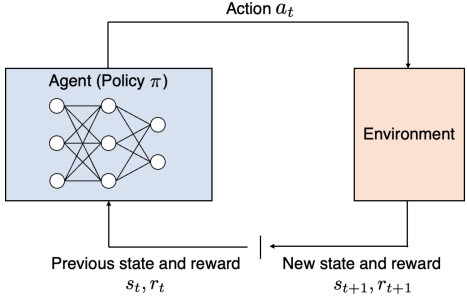

Fig. 3. The learning loop of RL algorithms [154].

In particular, block coordinate descent (BCD), which is a specific form of AO, has been extensively employed in the design of wireless systems. BCD is an iterative optimization algorithm utilized to solve problems that are separable into blocks, or groups, of variables. It is a variant of the coordinate descent (CD) method [152], where, instead of optimizing over a single variable at a time, one optimizes over a group of variables (or "block" of variables) at each iteration. Block cyclic coordinate descent (BCCD) is a special case of BCD where blocks are updated in a cyclic order. Additionally, inexact BCD is a variant of the BCD algorithm, where each block subproblem is solved approximately at every iteration, yet sufficiently well to ensure the algorithm's overall convergence. [153].

*5) Deep Reinforcement Learning:* Reinforcement learning (RL) is a type of machine learning in which an agent learns to make decisions by interacting with its environment as illustrated in Fig. 3. In the RL setup, an autonomous agent observes a state $s_t$ from its environment at timestep $t$, subsequently interacting with the environment by taking an action $a_t$. When the agent takes an action, the environment and the agent transition to a new state $s_{t+1}$ based on the current state $s_t$ and the action taken $a_t$. The sequence of actions that maximizes the agent's performance is determined by the rewards provided by the environment. It is noteworthy that with each transition of the environment to a new state $s_{t+1}$, a scalar reward $r_{t+1}$ is provided to the agent as feedback. The objective of the agent is to learn a policy $\pi$ that maximizes the cumulative reward over time. The policy, in essence, dictates the course of action to be taken in response to a given state. An optimal policy is defined as any policy that maximizes the expected return within the environment.

Deep reinforcement learning (DRL) is an integration of RL and deep learning, wherein deep neural networks are employed to approximate complex policies and value functions [154]–[156]. DRL has achieved notable scccess in numerous research domains, including wireless communication [157], [158], e.g., RIS optimization [159]–[163], and UAV trajectory design [164]–[168]. DRL algorithms may be classified into two categories: value-based DRL and policy-based DRL [154]. In value-based DRL, DRL algorithms attempt to approximate the value function, which is the estimation of the expected long-term reward of a state. Notable value-based DRL algorithms include deep Q-network (DQN) [169] and double DQN [170]. In contrast, policy-based DRL algorithms target the modeling of the policy and typically use deep neural networks to learn policies. Famous policy-based DRL algorithms include REINFORCE [171], trust region policy optimization (TRPO) [172], and proximal policy optimization (PPO) algorithm [173]. There is also a hybrid, called actor-critic approach, which employs both value functions and policy search [174]. The actor-critic architecture has been widely adopted as the foundational framework of subsequent algorithms, including asynchronous advantage actor-critic (A3C) [175], deep deterministic policy gradient (DDPG) [176], and soft actor-critic (SAC) [177].

*6) Others:* In what follows, we provide a brief overview of the other optimization techniques that have been employed in the reviewed literature.

- **Taylor Series Expansion:** The Taylor series expansion is a mathematical technique that approximates a function as an infinite sum of terms calculated from the function's derivatives at a single point. This method is frequently employed to approximate complex objective functions and constraints, thereby facilitating the optimization process. Specifically, the FOTE, which utilizes only the first derivative, and the second-order Taylor expansion (SOTE), which incorporates up to the second derivative term, are frequently employed in the literature.

- **S-procedure [178]:** The S-Procedure is a useful tool for reformulating a non-convex problem as a convex SDP problem. In essence, the S-procedure lemma facilitates the conversion of a quadratic inequality into a more manageable form, such as a set of linear matrix inequalities (LMIs), which can then be employed in optimization problems.

- **Schur Complement [179]:** The Schur complement is a technique used in linear algebra to reduce a block matrix to a simpler form and analyze positive definiteness, invertibility, or solve systems of equations. The Schur complement is frequently employed to transform nonlinear matrix inequalities into linear matrix inequalities, thereby rendering them manageable in the context of convex optimization.

- **Bernstein-Type Inequality [180]:** Bernstein-type inequalitys (BTIs) are probabilistic inequalities that bound the probability of a sum of independent random variables deviating from its expected value. Notable instances of BTIs include the well-known Markov inequality, the Chebyshev inequality, and the Chernoff bounds [181].

- **Sign-Definiteness Lemma [182]:** The sign-definiteness lemma is a tool for analyzing the nature of a matrix

or function. For optimization problems with many constraints, it facilitates the transformation of such semi-infinite problems into more straightforward LMIs. The sign-definiteness lemma has been generalized to the case of complex-valued matrices and multiple norm-bounded uncertainties in [183].

- **Lagrange Duality Method [121]:** The Lagrange duality (LD) method transforms a constrained optimization problem into a dual problem, which may prove to be more manageable. By introducing Lagrange multipliers and constructing the Lagrangian, we can identify the dual function and formulate the dual problem. In many cases, the dual problem provides a lower bound on the primal problem. In cases where the primal problem is convex, strong duality holds, meaning that solving the dual problem yields the same optimal value as the primal problem.

- **Penalty Method [184]:** The penalty-based method is a technique used to solve optimization problems by adding a penalty term to the objective function. This approach serves to eliminate solutions that do not align with the specified criteria. These methods are particularly useful when one wants to enforce constraints or incorporate regularization in the optimization process. The augmented Lagrangian method (ALM) [185] integrates the concepts of penalty methods and Lagrange multipliers to enforce constraints more effectively and stably than individual methods for constrained nonlinear optimization problems.

- **Alternating Direction Method of Multipliers [186]:** Alternating direction method of multipliers (ADMM) is an iterative optimization algorithm designed to solve complex problems by dividing them into smaller subproblems that can be optimized more easily and often in parallel. It introduces Lagrange multipliers (dual variables) to handle constraints and adds a penalty term to improve convergence, similar to ALM. This method is particularly effective for large-scale and distributed optimization problems.

- **Projected Gradient Descent [187], [188]:** Projected gradient descent (PGD) is a variant of the standard gradient descent method employed to solve constrained optimization problems, where the solution is required to satisfy certain constraints at each step. The fundamental premise of PGD is the integration of gradient descent with a projection step, ensuring that the iterates remain within a feasible region, thereby satisfying the constraints of the problem. PGD is particularly useful when dealing with constrained optimization problems where constraints cannot be simply handled by penalizing the objective (e.g., using a Lagrange multiplier or penalty method).

- **Manifold Optimization [189]–[192]:** Manifold optimizations (MOs) naturally address rank and orthogonality constraints by considering the problem as an optimization over a smooth manifold. The optimization over a manifold is locally analogous to that in the Euclidean space, and thus a well-developed gradient descent can find its counterpart on manifolds. In particular, the Riemannian conjugate gradient (RCG) method is an extension of the standard conjugate gradient method to manifolds. The Manopt toolbox [193] is a software dedicated to simplifying experimentation with state-of-the-art Riemannian optimization algorithms.

- **Rank-One Constraint Relaxation:** One possible approach to optimization problems with rank-one constraints is rank-one constraint relaxation. This method involves first solving the relaxed problem without the rank-one constraint and then constructing a rank-one solution based on the solution of the relaxed problem if its solution is not rank-one [118]. Alternatively, rank minimization (RM) can be employed, whereby the rank function is first approximated by a tractable function, and then the approximate rank function is added to the original objective function as a penalty term [194]. Sequential rank-one constraint relaxation (SROCR) [195] is another method to gradually relax the rank-one constraint, thereby facilitating the identification of a feasible solution.

- **Benders Decomposition [196]:** Benders decomposition is an algorithm used to solve large-scale optimization problems that have a specific structure, typically involving complicating variables or coupled constraints that make the problem hard to solve directly. It decomposes a complex problem into two constituent parts: a master problem and a subproblem. The subproblems are then used to generate cuts (constraints) that are added iteratively to the master problem, gradually refining its feasible region. A generalized version of the original Benders decomposition, known as generalized Benders decomposition (GBD), has been developed to accommodate nonlinear and non-convex settings [197].

- **Polyblock Approximation [198]:** Polyblock approximation is an algorithmic technique used to solve certain classes of non-convex optimization problems, particularly monotonic optimization problems. It does so by using a collection of "polyblocks", which are structured geometric objects made of axis-aligned boxes, to approximate the feasible set or upper-bound of the objective.

- **Inner Approximation [199]:** Inner approximation is a technique in which a complex feasible set or objective is approximated using a simpler or smaller subset that is guaranteed to be feasible. The objective is to solve a simpler version of the original problem, where all candidate solutions are guaranteed to be feasible, though possibly suboptimal.

- **Simultaneous Perturbation Stochastic Approximation [200]:** Simultaneous perturbation stochastic approximation (SPSA) is a gradient-free optimization algorithm used to solve optimization problems where the objective function is black-box, or expensive to evaluate. It is particularly effective in high-dimensional problems, where the cost of computing or estimating gradients would be prohibitive.

- **Big-M Formulation [201]:** The Big-M formulation is a mathematical technique used to model logical constraints, such as those involving binary variables. It is commonly used in mixed-integer linear programming (MILP) to encode conditional relationships using binary variables and large constants.

- **Arithmetic-Geometric Mean (AM-GM) Inequality:** The arithmetic-geometric mean inequality states that for any set of non-negative real numbers, the arithmetic mean (AM) is always greater than or equal to the geometric mean (GM), with equality holding if and only if all the numbers are equal.
- **Fixed-Point Theorem [202]:** A fixed-point theorem is a mathematical result that guarantees the existence of a point that remains unchanged under a given function. There are several types of fixed-point theorems; however, the Banach fixed-point theorem is of particular importance due to its role in ensuring the convergence of iterative methods, such as successive approximations, to a unique solution in the context of numerical equation solving.

## III. Survey of ISAC Systems for PLS

This section presents a comprehensive review of the literature on PLS for ISAC systems, categorizing the existing papers based on the scenarios addressed.

### A. Generic ISAC Systems

This subsection examines literature on generic ISAC systems, which consist of an ISAC BS, communication users, and sensing targets, as depicted in Fig. 2. Table I and Table II present a summary of these papers.

*1) Single-User Scenario:* The design of transmit BF in multi-antenna ISAC BSs has been a subject of extensive research in the literature. One of the initial works on secure ISAC systems [203] addressed three issues concerning a single target (eavesdropper): maximizing the SR, maximizing radar SINR, and minimizing transmit power. The Taylor series approximation was employed for the SR function to address the non-convexity of the optimization problems. The authors of [204] addressed the scenario of multiple eavesdroppers and a distinct sensing target in a subsequent study. They suggested an AN-assisted BF approach that was designed to maximize the worst-case SR while adhering to constraints on the sensing SNR and power budget. In contrast to the Taylor series approximation employed in [203], they adopted a first-order approximation of the objective function and developed an AO algorithm to identify the optimal solution. The authors of [205] formulated a problem analogous to that in [204], yet they showed that the problem can be reformulated as a convex SDP problem and proposed two optimal semiclosed-form BF solutions and one suboptimal closed-form solution. In [206], the authors considered a DFRC system in vehicle-to-vehicle (V2V) networks. The authors optimized AN-aided BF to maximize the SR while adhering to radar SINR and transmit power constraints, consistent with previous studies. They proposed an inexact BCD algorithm based on SDP to tackle the non-convex problem. They optimized AN-aided BF to maximize the SR while satisfying radar SINR and transmit power constraints, similar to the above works. To address the non-convex problem, they proposed an inexact BCD algorithm based on SDP.

Conversely, despite its superior capacity, the hardware cost and complexity of digital BF are substantial, and antenna selection emerges as a cost-effective and efficient alternative to fully digital BF [233]. In this regard, the authors of [209] proposed joint antenna allocation and transmit BF for a secure ISAC system. They proposed an AO-based solution to address the BPG problem, which is subject to the SR constraint. Furthermore, in [207], the authors considered beamforcing design in near-field ISAC systems with extremely large-scale (XL)-MIMO, with the objective being to maximize the SR while satisfying the constraints on a BPG and power budget. To this end, they developed a two-stage optimal solution and a zero-forcing (ZF)-based suboptimal solution to the formulated problem.

Despite the assumption of perfect CSI at the ISAC BS in the aforementioned papers, attaining perfect CSI for legitimate users and eavesdroppers is challenging in practice. To address this issue, the authors of [85] considered two types of CSI error models for eavesdroppers, namely bounded and Gaussian CSI errors. For both scenarios, they formulated the robust BF design problem as the BPG problem under the constraints of worst-case SR and SOP, respectively. Also, the authors of [212] proposed hybrid BF in the context of imperfect CSI with the bounded CSI model. Furthermore, the authors of [216] considered a scenario in which the BS serves as one legitimate receiver and multiple energy receivers that are considered as eavesdroppers. In this scenario, the authors proposed secure BF design to maximize the SR under the energy-harvesting constraint and the sensing SINR constraints for both perfect and imperfect CSI of the eavesdroppers. In addition to DFRC, robust BF design has also been studied for a radar and communication coexisting (RCCE) system in [214] under the norm-bounded CSI error model [234]. Furthermore, instead of the instantaneous CSI, the authors of [215] considered the statistical CSI of both the user and the eavesdropper, and proposed a transmit BF design to maximize SR.

Meanwhile, the authors of [208] proposed a two-stage transmission protocol including the sensing of the eavesdropper's CSI and secure communication to a legitimate receiver. To this end, they derived the closed-form CRB to describe the relationship between the estimated angle range of the eavesdropper's location and the number of sensing beams, and designed the transmit BF to minimize the eavesdropper's rate while satisfying the requirement for the communication user's rate. More recently, pilot-based codebook artificial noise [235], in which the exchanged pilot signals between the transceiver serve as an encryption key, has been integrated into an ISAC system [213] as a means to enhance PLS without the eavesdropper's CSI.

In [116], the authors addressed the deterministic-random trade-off in the dual-functional ISAC waveform [236] and developed a secure ISAC system with random signaling. The deterministic-random trade-off indicates the fundamental trade-off between sensing and communication performances. More specifically, the entropy of the ISAC waveform should be high to convey more information, while deterministic signals are preferred for sensing. In [116], the authors introduced EMMSE as the sensing performance metric and optimized it

TABLE I
Summary of papers on generic ISAC systems (single-user).

| Reference | User(s) | Eve(s) | Target(s) | KPIs (Com. / Sen.) | Eve's CSI | Opt. Variables | Optimization Techniques |
|-----------|---------|--------|-----------|--------------------|-----------|----------------|------------------------|
| [203] | S | S | S (Eve) | SR / SINR | Perfect | BF | AO, FOTE |
| [204] | S | M | S | SR / SNR | Perfect | BF, AN | AO, First-Order Approx. |
| [205] | S | M | S | SR / SNR | Perfect | BF | LD, SDR |
| [206] | S | M | M (Eve) | SR / SINR | Perfect | BF, AN | Inexact BCD, LD, SDP |
| [207] | S | M | M (Eve) | SR / BPG | Perfect | Beam-Focusing, Radar Waveform | FP, Line Search, SDR |
| [85] | S | M | M (Eve) | SR or SOP / BPM and CC | Imperfect | BF, AN | Line Search, SDR, S-procedure |
| [208] | S | S | S (Eve) | Eve's Rate / CRB | Unknown | # of Sensing Beams, BF | AO, Fixed-Point Theorem, LD |
| [209] | S | M | M | SR / BPM | Perfect | Antenna Allocation, BF | AO, Penalized SCP [210], SDR |
| [116] | S | S | S | SR / EMMSE | Perfect | Waveform | FOTE, Iterative RM [211], SDP |
| [212] | S | S | S | SR / BPM | Perfect / Imperfect | Analog / Digital BF | SCA, SDR, S-procedure |
| [213] | S | S | S (Eve) | SR / CRB | Unknown | BF, PCAN | AO, SDR |
| [214] | S | S | S (Eve) | SR / INR | Imperfect | BF, Radar Waveform | FP, Line Search, SDR, S-Procedure |
| [215] | S | S | S (Eve) | SR / BPM | Statistical | BF | FOTE, Large-System Analysis |
| [216] | S | M | M (Eve) | SR / SINR | Perfect / Imperfect | BF | FOTE, SDR, S-procedure |

S: Single, M: Multiple.

under the constraints of SR and power budget.

*2) Multi-User Scenario:* In the following, we proceed to deliberate on three additional scenarios: the presence of a single eavesdropper and target, the presence of a single eavesdropper or target, and the presence of multiple eavesdroppers and targets.

*Single Eavesdropper and Target:* The papers [217]–[224] considered a scenario with a single eavesdropper and a single sensing target that can act as a potential eavesdropper. In this scenario, the papers [218], [219], [220] proposed AN-aided secure BF. Specifically, the authors in [218] and [219] sought to minimize the eavesdropper's SINR under the constraints of transmit power budget and MSE between the designed and the desired beampattern. In contrast, the authors in [220] aimed to maximize the communication user's rate under the constraints of the eavesdropper's SINR, MSE between the designed and the desired beampattern, and power budget. In addition, the authors of [218] proposed robust BF with imperfect CSI and target direction uncertainty. Furthermore, the authors of [222] also proposed robust BF in low-earth orbit (LEO) satellite systems to maximize the worst-case SR while satisfying the sensing SNR constraint. Contrary to digital BF, the authors of [217] proposed a hybrid BF approach as a means of reducing hardware costs and power consumption. They formulated a secure BF design under imperfect CSI of the eavesdropper and target location as a maximization of the minimum user's rate, subject to constraints of the radar SINR, eavesdropper's rate, and power budget.

Meanwhile, the authors of [223] proposed a two-phase BF design under their channel uncertainty model of the eavesdropper's channel. In the initial phase, the location of a potential eavesdropper is estimated, and in the subsequent phase, BS employs AN-aided BF based on the estimated information in the second phase. Specifically, they attempted to minimize the CRB by designing the sensing waveform in the first phase, and then maximize the system sum rate under the SR constraint by optimizing transmit BF. On the other hand, the authors of [221] introduced a novel sensing-specific imperfect CSI model, integrating the sensing capacity with the norm bound of the CSI uncertainty. Under the CSI error model, they proposed to maximize the SR under the CRB constraint. To address the non-convexity of the original problem, they developed a BCD-based iterative algorithm, leveraging SDR and SCA techniques.

In addition, the authors of [224] considered a scenario in which the target's location is unknown and random, while its distribution information is available a priori. In this scenario, they derived PCRB [91] and its tight approximation. Then, they formulated the optimization problem of AN-aided transmit BF to maximize the worst-case SR among all possible target locations, subject to a maximal threshold on the sensing PCRB. By employing the Charnes-Cooper equivalent transformation [147] and SDR techniques, they transformed the original non-convex problem into a convex form. They then proposed an optimal solution and two suboptimal solutions with low complexity.

*Single Eavesdropper or Target:* The authors in [79], [225] considered a scenario with a single eavesdropper and multiple targets. In [225], the authors assumed that the CSI of the eavesdropper's channel was unavailable and introduced *long-term* eavesdropping SINR as a security metric. Based on this metric, the authors considered the sum rate maximization problem and the jamming power maximization problem. In contrast, the authors of [79] proposed a deep learning approach to optimize transmit BF that does not require explicit CSI. In this approach, as illustrated in Fig. 4, multi-layer perceptrons (MLPs) take pilots and echo signals as input and output a BF matrix. Furthermore, they proposed a loss function for the training derived from first-order optimality conditions and demonstrated by simulations that their method outperforms an

TABLE II
SUMMARY OF PAPERS ON GENERIC ISAC SYSTEMS (MULTI-USER).

| Reference | User(s) | Eve(s) | Target(s) | KPIs (Com. / Sen.) | Eve's CSI | Opt. Variables | Opt. Techniques |
|---|---|---|---|---|---|---|---|
| [217] | M | S | S | User's Rate / SINR | Imperfect | Analog / Digital BF, Receive filter | BCD, MM |
| [218] | M | S | S (Eve) | SINR / BPM | Perfect / Imperfect | BF, AN | FP, SDR |
| [219] | M | S | S (Eve) | SINR / BPM | Perfect | BF and AN | FP, SDR |
| [220] | M | S | S (Eve) | SINR / BPM | Perfect | BF and AN | SDR, Weighted MMSE |
| [221] | M | S | S (Eve) | SR / CRB | Imperfect | BF | BCD, SCA, SDR, S-procedure |
| [222] | M | S | S (Eve) | SR / SNR | Angle Uncertainty | BF, Receive Filter | AO, CCCP, FOTE |
| [223] | M | S | S (Eve) | SR / CRB | Imperfect | Time Allocation, Sensing Waveform, BF, AN | Polyblock Approx., SDP, S-procedure |
| [224] | M | S | S (Eve) | SR / PCRB | Statistical | BF, AN | FP, SDR, Schur Complement |
| [225] | M | S | M | SINR / BPM | Unknown | BF | SDR, ZF |
| [79] | M | S | M | Eve's SINR / TIP | Unknown | BF | Deep Learning (MLP) |
| [80] | M | M | S | Eve's SINR / SINR | Perfect / Unknown | BF, Radar Receive Filter | BCD, FP, SDR |
| [226] | M | M | S | SINR / CRB | Perfect | BF | SDR |
| [227] | M | M | S (Eve) | SR / BPM | Perfect | BF and AN | BCD, SDR |
| [71] | M | M | M (Eve) | SINR / BPM and CC | Perfect / Imperfect | BF | SDR, ZF |
| [228] | M | M | M (Eve) | SR / BPM | Imperfect | snapshot length, BF, AN | BCD, Inner Approx., SDR |
| [70] | M | M | M (Eve) | SOP / BPM | Imperfect | BF, AN | BTI, SDR |
| [229] | M | M | M | User's rate / - | Imperfect | BF, Jamming | Schur Complement, Sign-Definiteness, S-procedure |
| [230] | M | M | M (Eve) | SINR / MSE | Perfect | BF | FP, SDR |
| [31] | M | M | M (Eve) | SR / CRB | Unknown | BF, AN | FP |
| [231] | M | M | M (Eve) | SR / CRB | Perfect | Subcarrier and Power Allocation, Power Splitting Ratio | LD |
| [232] | M | M | D | SINR / BPM and CC | Statistical / Unknown | BF | SDR, ZF |

S: Single, M: Multiple, D: Double.

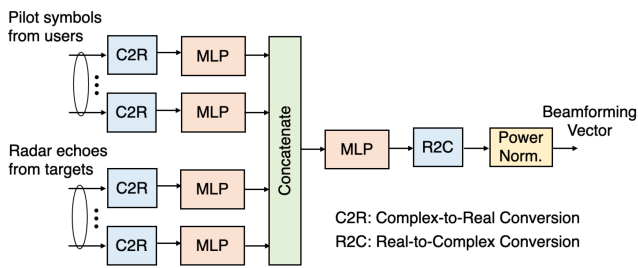

Fig. 4. Deep learning architecture for BF optimization in [79].

SDP-based method.

Meanwhile, the authors of [71], [80], [226], [227] considered a scenario comprising multiple eavesdroppers and a single target. The authors of [71], [80] consider two scenarios, one with and one without eavesdropper's CSI. In scenarios where eavesdropper's CSI is accessible, the authors proposed a joint optimization of transmit BF and the radar receive filter, aimed at minimizing the maximum eavesdropper's SINR. This optimization was undertaken to ensure satisfaction of the quality of service (QoS) requirements of legitimate users, the radar output SINR constraint, and the total transmit power budget. Conversely, when eavesdropper's CSI is not available, they proposed to maximize the AN power while maintaining the same constraints.

In [226], the authors formulated the transmit BF design problem as a transmit power minimization under the constraints on SINRs of legitimate users and eavesdroppers, as well as CRB. They addressed this problem by employing an SDR technique with a slight modification, proving that the SDR applied to their problem is always tight. In contrast, the authors in [227] formulated the transmit BF design as a maximization of the sum SR, which was addressed by employing SDR and BCD methods.

*Multiple Eavesdroppers and Targets:* A number of papers have examined a scenario involving multiple sensing targets, which can be considered potential eavesdroppers [31], [70], [71], [228]–[232], [237]. The authors of [228] proposed a novel optimization framework, where the available resources are jointly optimized over a sequence of variable-length snapshots[5]. Specifically, they jointly optimized the duration of each snapshot and AN-aided BF for maximizing the sum SR, accounting for imperfect CSI of both legitimate users and the eavesdroppers. The authors of [70] also took into account the imperfect CSI of both legitimate users and eavesdroppers. They formulated the robust transmit BF design problem as the BPG problem under the constraints on outage probabilities for the legitimate users and the eavesdropper. In [71], the authors addressed the transmit BF design problem, aiming to minimize a sensing performance metric. This metric comprises BPG error and mean squared CC, constrained by user's and eavesdropper's SINRs. They proposed an optimal solution based on an SDR, along with a low-complexity design based on ZF, assuming perfect CSI. Additionally, they presented a robust design with imperfect CSI for legitimate users and angle uncertainty for targets. Recently, in [232], the authors considered a scenario in which active and passive eavesdroppers collude to intercept the confidential information. In this scenario, they proposed a BF design to optimize the same sensing metric as in [71], subject to SINR requirements of the communication users and the active eavesdropper, the minimum outage requirement for the passive eavesdropper, and power budget.

[5]The transmission of the actual beam patterns is referred to as one snapshot in ISAC systems [228].

Meanwhile, the authors of [231] considered OFDM signals for ISAC and optimized the resource allocation to maximize the SR under the CRB constraint. In [229], the authors proposed a sensing-assisted communication and jamming protocol, and they designed communication and jamming BF to minimize minimize the transmit power under the constraints of achievable rates of legitimate and suspicious receivers. In addition, in [230], the transmit BF design was considered for two types of receivers: conventional and sensing-dedicated receivers. For the conventional receiver, which is incapable of extracting communication signals from the ISAC signal, the problem was formulated as a minimization of eavesdropper's SINRs. For the ISAC-dedicated receiver, which can extract communication signals from the ISAC signal, the problem was formulated as a minimization of the BPG error.

Furthermore, the authors of [31], [237] proposed a two-stage method for secure ISAC as shown in Fig. 5. In this method, the ISAC BS first emits an omnidirectional waveform to search for potential eavesdroppers by employing the combined Capon and approximate maximum likelihood technique (Fig. 5a)[6]. Then, SR expressions are formulated, which are a function of the eavesdropper's sensing accuracy. The problem was formulated as a weighted optimization, with the objective being to maximize the SR while minimizing the CRB for target sensing with the aid of AN (Fig. 5b). An iterative algorithm is developed to address the problem, where the sensing and security functionalities provide mutual benefits at each iteration by improving the estimation accuracy.

*Summary:* A number of papers addressed transmit BF design in various scenarios based on different key performance indicators (KPIs). Several papers tackled the issue of CSI and target direction uncertainties, which is important in practice. In this case, deep learning-based BF design, as in [79], appears to be promising. Also, AN-aided BF has been widely adopted in the literature, which is effective when the eavesdropper's CSI is imperfect or unavailable. Furthermore, the two-stage framework, proposed in [31], [208], [223], is a particularly promising approach when prior information regarding the location of the eavesdropper is unavailable, as it offers a multifaceted benefit by leveraging the strengths of both sensing and PLS functionalities. Meanwhile, the randomness

---

[6]It is assumed that the locations of all communication users are accessible at the BS, thereby enabling the estimation of eavesdropper locations through the subtraction of these users' locations from the reflected echo

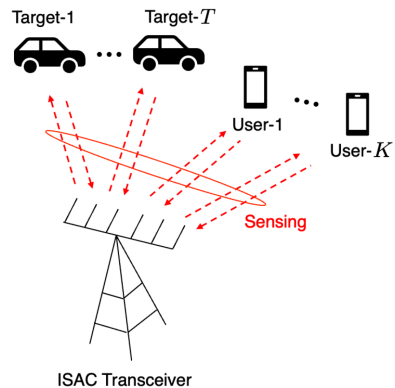

(a) First step: sensing with omnidirectional waveform.

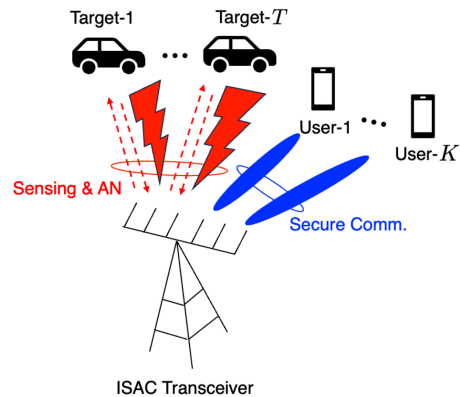

(b) Next step: AN-aided secure BF and sensing.

Fig. 5. Two-stage method for secure communication and sensing in [31].

of sensing signals is often overlooked in the literature. The interplay between sensing and communication in secure ISAC systems, characterized by the deterministic-random tradeoff [116], is thus a crucial subject to be explored in future studies.

### B. Cell-Free Network

Cell-free (CF) MIMO systems possess a substantial number of access points (APs) that are individually controllable and distributed over a vast area, with the capacity to concurrently serve a limited number of users. This configuration offers advantages over traditional cellular architectures, including improved spectral efficiency, energy efficiency, uniform service quality, and enhanced robustness to interference and shadowing effects [243]–[247]. Consequently, the CF-ISAC architecture, as illustrated in Fig. 6, offers substantial advantages over single-BS ISAC architecture with regard to both sensing and communication performances [248], [249]. Recently, several papers studied PLS techniques for CF-ISAC, which are summarized in Table III.

The authors of [238] considered a cell-free MIMO network equipped with ISAC APs. They proposed a novel AN-based ISAC waveform model and formulated its optimization problem as CRB minimization under the constraints of the user's SINR and eavesdropper's SNR. In contrast, the authors

TABLE III
SUMMARY OF PAPERS ON CELL-FREE NETWORK.

| Reference | User(s) | Eve(s) | Target(s) | KPIs (Com. / Sen.) | Eve's CSI | Opt. Variables | Optimization Techniques |
|---|---|---|---|---|---|---|---|
| [238] | M | S | S (Eve) | SINR / CRB | Perfect | BF, AN | SDR |
| [239] | M | S | S (Eve) | SR / SINR | Perfect | BF, Transmit / Receive Quantizers | FOTE, MM, SDR |
| [240] | M | - | S | SINR / SINR | Perfect | - | - |
| [241] | M | M | S | SINR / TDP and SNR | Perfect | BF, Radar Waveform | SDR |
| [242] | M | M | S | SR / SNR | Perfect | BF | Inner Approx., SCA |

S: Single, M: Multiple.

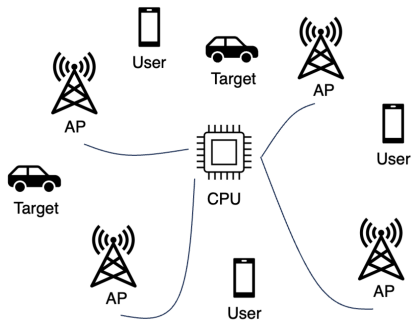

Fig. 6. A CF-ISAC scenario.

of [242] formulated the joint design of information and sensing BF as a sensing SNR maximization problem under the constraint of the SR and transmit power budget. Furthermore, the authors of [241] considered a CF ISAC system in the presence of both information and sensing eavesdroppers. They optimized information and sensing BF to maximize the TDP while ensuring the minimum SINR requirements for communication users, the maximum tolerable SNR constraints for information eavesdroppers, and the maximum detection probability constraints for sensing eavesdroppers. While the aforementioned works assume ideal fronthaul links, the authors in [239] considered a CF-MIMO with limited fronthaul links, and they optimized the transmit BF and transmit/receive quantizers to maximize the sensing SINR under the constraints on SR, fronthaul rate, and power budget.

In addition, the authors of [240] considered a scenario in which a cell-free massive MIMO architecture is exploited by malicious actors. These actors utilize multiple APs to illegally sense an aerial legitimate target while communicating with users, one of which is suspected of illegal activities. The authors proposed an anti-malicious strategy based on proactive monitoring. This strategy involves the monitor sending jamming signals to the suspicious user to intercept the communication and to the legitimate target to hinder the malicious sensing attempts.

*Summary:* A CF-MIMO ISAC topology exhibits several advantages over the single-cell ISAC, including larger monitoring areas, broader sensing coverage, and a wider range of sensing angles [249]. On the other hand, there are several intriguing extensions for future research in secure CF ISAC systems. Initially, the aforementioned works can be extended to scenarios with practical constraints, such as imperfect CSI and time synchronization among ISAC transmitters. Addition-ally, exploring the potential of CF-MIMO ISAC in scenarios with multiple sensing targets offers a promising avenue for future research.

### C. Full-Duplex ISAC

Recent studies have demonstrated that FD ISAC systems exhibit substantial enhancements in terms of power and spectral efficiency when compared to their half-duplex (HD) counterparts [134], [255], [256]. However, it should be noted that the majority of existing literature on PLS for HD ISAC systems may not be directly applicable to FD ISAC systems. This is primarily due to the significant self-interference caused by both the communication signal and radar echo. To address this issue, the literature has explored PLS techniques for FD ISAC systems, as summarized in Table IV.

*1) Single-User Scenario:* In [250], the authors proposed a two-step approach to secure uplink (UL) communication with a single user in an ISAC system. Specifically, in the first phase, the BS synthesizes a wide beam to localize and jam an aerial eavesdropper under the SR constraint, and in the second phase, the BS maximizes the SINR of the communication user while satisfying the constraints of the eavesdropper's SINR and radar echo SNR. They proposed a joint optimization of the radar waveform and receive BF, employing AO and SCA-based algorithms.

Contrary to [250], which focused on uplink communication and downlink sensing, the authors of [251], [257] addressed both UL and downlink (DL) communications with a single UL/DL user for the FD ISAC system. Specifically, they developed an iterative algorithm to optimize the transmit BF, the radar waveform, the communication receive filter, and the radar receive filter by developing an iterative algorithm. The authors formulated the problems of minimizing the eavesdropper's SINR and maximizing SR, subject to the sensing and the communication constraints. The simulation results demonstrated that their scheme approaches the performance of a system that designs secure communication and sensing separately.

*2) Multi-User Scenario:* A multi-user scenario for FD-ISAC is depicted in Fig. 7, where multiple malicious targets attempt to intrude on UL and DL communication between legitimate UL/DL users and a BS. In this setting, the authors of [252] considered a secure FD ISAC system that exploits AN at the DFRC BS, in conjunction with UL/DL BF design and UL power allocation. They addressed the transmit power minimization problem, subject to the constraints on UL/DL SRs and radar ISMR. In a similar context, the authors of

TABLE IV
SUMMARY OF PAPERS ON FD ISAC.

| Reference | User(s) | Eve(s) | Target(s) | KPIs (Com. / Sen.) | Eve's CSI | Opt. Variables | Optimization Techniques |
|-----------|---------|--------|-----------|--------------------|-----------|----------------|-------------------------|
| [250] | S | S | S (Eve) | SINR / SNR | Imperfect | Radar Waveform, Receive BF | AO, FOTE, SCA |
| [251] | D | S | S (Eve) | SR / SINR | Perfect / Imperfect | BF, Radar Waveform, Communication / Radar Receive Filter | AO, Bisection Search, FP, SCA, SDR, S-procedure |
| [252] | M | M | M (Eve) | SR / ISMR | Perfect | UL Power Allocation, DL BF, AN | SCA |
| [253] | M | S | M | SINR / SINR | Perfect | Transmit / Receive BF, AN, Radar Receive Filter | FOTE, FP, SCA, SDR |
| [254] | M | M | M (Eve) | SR / ISMR | Perfect | UL Power Allocation, DL BF, AN | FOTE, BCCD |

S: Single, M: Multiple, D: Double.

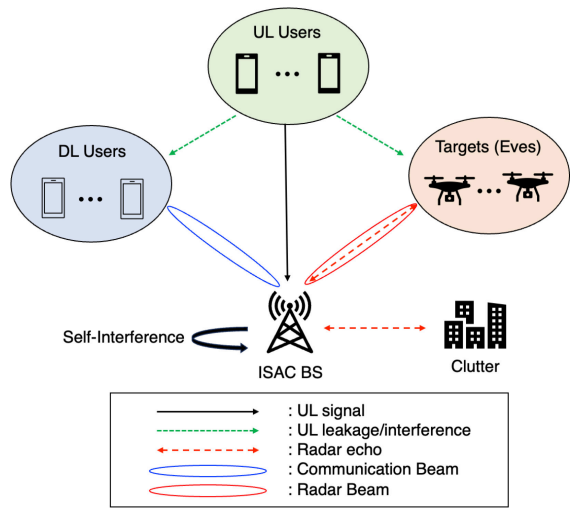

Fig. 7. A FD-ISAC scenario considered in [252], [254].

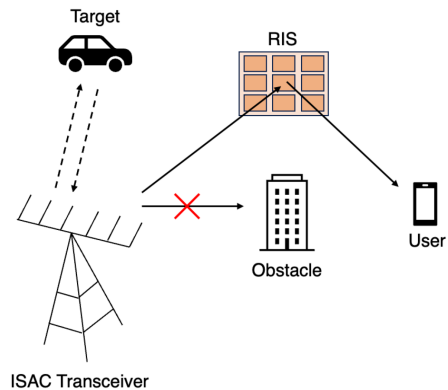

Fig. 8. A RIS-assisted ISAC network.

[254] attempted to maximize the UL/DL sum SR under various power budgets and an ISMR constraint. To address this problem, they developed an iterative joint Taylor-block cyclic coordinate descent method, which alternates between two subproblems. One of these subproblems yields UL BF in closed form, while the other approximates the solution for UL power allocation, AN covariance, and DL BF vectors.

Meanwhile, the authors of [253] investigated the energy efficiency of the secure ISAC system, which is defined as the ratio of the total DL SR to total power consumption at the BS. They proposed a joint optimization of transmit and receive BF, along with AN, to achieve maximum energy efficiency under the constraints of the user's and target's SINR. This problem was addressed through the utilization of SDR, SCA, and FP techniques.

*Summary:* The aforementioned works have revealed that joint transceiver design schemes are effective in addressing the PLS problem for FD ISAC and have demonstrated their effectiveness in comparison to HD ISAC. However, these papers have made several ideal assumptions to simplify the design and analysis. For example, they have assumed that perfect CSI is available and that hardware impairment is ignored. It is of both theoretical and practical interest to investigate the impact of relaxing these assumptions on the secure FD ISAC system.

### D. RIS-Assisted ISAC

*1) Passive RIS-Assisted ISAC:* Recent attention has been given to RIS due to its potential to intelligently reconfigure wireless communication environments with cost-effective reflecting elements [282]–[284]. Specifically, based on electromagnetic (EM) scattering principles, each reflecting element is able to reflect the incident radio frequency signal with an adjustable reflecting coefficient (amplitude and phase). Through the strategic design of these coefficients, RIS can enhance signal reception at desired destinations and/or suppress interference to unintended users, thereby creating favorable propagation conditions.

As illustrated in Fig. 8, a RIS-assisted ISAC network employs virtual LOS links between the BS and the communication user/target by passively reflecting the impinging radio signals. The effectiveness of this approach is particularly evident when the LOS link between the BS and the communication user/target is obstructed by obstacles, such as buildings. The intelligent deployment and design of RIS holds great potential for enhancing the received communication and radar performance.

Recently, due to the potential of RIS to enhance both communication and sensing performance, RIS-assisted ISAC systems have been the focus of extensive research [285], [286].

TABLE V
SUMMARY OF PAPERS ON PASSIVE RIS-ASSISTED ISAC (SINGLE-USER).

| Reference | User(s) | Eve(s) | Target(s) | KPIs (Com. / Sen.) | Eve's CSI | Opt. Variables | Optimization Techniques |
|-----------|---------|--------|-----------|--------------------|-----------|----------------|-------------------------|
| [258] | S | S | S (Eve) | SR / SINR | Perfect | BF, AN, RIS Phase Shifts | BCD, FOTE, MM |
| [259] | S | S | S (Eve) | SR / TIP | Perfect | BF, RIS Phase Shifts, Transmit Power, Power Allocation | AO, RCG |
| [260] | S | S | S (Eve) | SR / TIP | Imperfect | BF, RIS Phase Shifts | BCD, FOTE, SDR, S-procedure |
| [261] | S | S | S (Eve) | SR / SNR | Perfect | BF, AN, RIS phase shits | AO, FP, LD, MM, SDR |
| [262] | S | S | S (Eve) | SR / CRB | Perfect | BF, Radar Waveform, RIS Phase Shifts | AO, FOTE, SCA, Schur Complement, SDR |
| [263] | S | S | S | SR / SNR | Unknown | Transmit / Receive BF, AN, RIS Phase Shifts | AO, SCA, SDR |
| [264] | S | S | S (Eve) | SR / SNR | Perfect / Imperfect | Power Allocation, BF, Radar Waveform, RIS Phase Shifts | AO, BCD, FP, RCG, SDR, S-procedure |
| [265] | S | M | M (Eve) | SR / TIP | Perfect | BF, RIS Phase Shifts | AO, DRL (SAC) |
| [266] | S | M | M (Eve) | SR / TDP | Perfect | BF, RIS Phase Shifts | BCD, FOTE, MM, SDR, Weighted MMSE |
| [267] | S | S | S (Eve) | SG / BPM | Perfect | Analog / Digital BF, RIS Phase Shifts | Bisection Method, Inexact BCD, LD, Penalty Method |

S: Single, M: Multiple, SG: Secrecy Gap.

TABLE VI
SUMMARY OF PAPERS ON PASSIVE RIS-ASSISTED ISAC (MULTI-USER).

| Reference | User(s) | Eve(s) | Target(s) | KPIs (Com. / Sen.) | Eve's CSI | Opt. Variables | Optimization Techniques |
|-----------|---------|--------|-----------|--------------------|-----------|----------------|-------------------------|
| [268] | M | S | S (Eve) | SR / BPM | Perfect | BF, AN, RIS Phase Shifts | AO, FP, LD, Penalty Method, RCG |
| [269] | M | S | S (Eve) | SINR / BPG | Perfect / Imperfect | BF, RIS Phase Shifts | AO, LD, MM, Penalty Method, SCA, Sign-Definiteness, S-Procedure |
| [270] | M | S | S (Eve) | SR / BPM | Perfect | BF, AN, RIS phase shifts | DRL (SAC) |
| [271] | M | S | S (Eve) | SR / TIP | Perfect | BF, RIS Phase Shifts | AO, FP, Local Search [272], SDR |
| [273] | M | S | S (Eve) | SINR / SNR | Perfect | BF, Radar Recei e Filter, RIS Phase Shifts | AO, FOTE, FP, MM, SDR |
| [274] | M | S | S | SINR / TDP | Perfect | BF, Radar Waveform, RIS Phase Shifts | AO, FP, SDR |
| [275] | M | S | S (Eve) | User's Rate / SNR | Perfect | BF, Radar Receive Filter, RIS Phase Shifts, Positions of MAs | AO, FOTE, FP, LD, MM, SOTE |
| [276] | M | S | S (Eve) | SR / SNR | Perfect | BF, AN, RIS Phase Shift, UAV Deployment | AO, SCA, SDR |
| [277] | M | M | M (Eve) | SR / SINR | Perfect | BF, RIS Phase Shifts | BCD, FOTE, SPSA |
| [278] | M | M | M (Eve) | SINR / BPG | Perfect | BF, AN, RIS Phase Shifts | AO, Penalty Method |
| [279] | M | M | M (Eve) | SR / SCNR | Perfect | BF, AN, RIS Phase Shifts | AO, ALM, Manopt toolbox |
| [280] | M | S | S (Eve) | SR / BPG | Imperfect | BF, RIS Phase Shifts | AO, FOTE, SCA, Schur Complement, Sign-Definiteness, SOC, S-Procedure |
| [281] | M | S | S (Eve) | SR / SNR | Perfect | BF, RIS Phase Shifts, User Association | ADMM, AO, FOTE, MM, Penalty Method, SCA, SDR |

S: Single, M: Multiple.

Additionally, RIS-ISAC systems have been examined for PLS problems. The existing literature on passive RIS-aided ISAC is outlined in Table V and Table VI, which are subsequently reviewed.

*Single-User Scenario:* RIS-assisted secure ISAC systems in the presence of a single communication user and a single sensing target have been studied in the literature. In [258], the authors proposed a joint design of the transmit BF and the RIS phase shifts to maximize the radar SINR under the SR constraint. To address the non-convex optimization problem, they proposed a BCD algorithm based on Taylor series approximation and MM methods. The authors of [259] also proposed joint design of the transmit BF and the RIS phase shifts, aiming to maximize the SR under the sensing metric, i.e., TIP. To this end, they proposed a RCG-based AO algorithm to address the formulated problem. Furthermore, the authors of [261], [287] proposed a joint design of transmit BF and RIS parameters for maximizing the SR under the radar SNR constraint. Assuming the perfect channel knowledge, they proposed a low-complexity AO algorithm based on the closed-form solution for RIS parameter updates that enables scalable RIS design. More recently, the authors of [267] proposed a hardware-efficient hybrid BF architecture for a RIS-assisted

ISAC system. The authors proposed a joint optimization of the analog/digital beamforming and RIS coefficients to maximize the communication secrecy gap, which is defined as the difference in SNR between a legitimate communication channel and an eavesdropping channel. To address the non-convex optimization problem, the authors developed a penalty dual decomposition framework.

In contrast to fully passive RIS, the authors of [262], [288], [289] examined the semi-passive RIS, which incorporates not only passive elements but also a dedicated sensor to detect the reflected signal from the target, thereby enhancing the sensing performance. Additionally, they considered two types of sensing targets (eavesdroppers): point and extended targets. In these settings, they optimized transmit BF and RIS phase shifts based on an AO-based optimization algorithm to simultaneously maximize the SR and minimize the CRB.

Despite the assumption of perfect CSI of the eavesdropper in the aforementioned works, the authors in [260] proposed robust BF and RIS design for imperfect CSI of the eavesdropper to maximize the SR. Furthermore, in [263], the authors considered the eavesdropper to be in a specific region with an unknown position and derived an approximate average SR. Subsequently, they developed an AO approach for designing transmit/receive BF and RIS phase shifts, with the objective of maximizing the SR. Meanwhile, the authors of [264] explored the application of RIS for both RCCE [290] and DFRC [56] systems. Assuming both perfect and imperfect CSI of the eavesdropper, they considered BF design to maximize the SR and demonstrated that RCCE can achieve a higher SR than DFRC.

Furthermore, the authors of [265] and [266] considered a scenario with multiple sensing targets (potential eavesdroppers). In [266], they considered a SR maximization problem under the constraint of TDP. More specifically, they used a weighted MMSE method to transform the original problem into a more tractable form and developed the BCD algorithm to solve the problem. On the other hand, the authors of [265] proposed a DRL-based approach to design transmit BF and RIS discrete phase shifts for maximizing the SR while ensuring sensing performance in terms of the TIP. Specifically, they employed a SAC for optimization of transmit BF and an AO algorithm for optimization of discrete phase shift of RIS.

*Multi-User Scenario:* A considerable amount of research has been dedicated to the study of RIS-assisted secure ISAC systems, with a particular focus on scenarios involving multiple communication users and a single sensing target. For instance, in the papers [268]–[271], [274], the authors explored the joint design of transmit BF and the RIS phase shifts in the context of a single sensing target. In particular, the authors of [268], [270] considered the SR maximization problem under the constraint of MSE between the desired and the designed beampatterns. The papers [268] and [270] addressed the problem by an AO algorithm with some techniques, such as the penalty-based method and RCG, and SAC, respectively. In contrast, the authors of [274] investigated the minimization of the eavesdropper's SINR under the constraints of TDP, QoS requirement for users, and power budget. The problem was addressed by an AO algorithm based on FP and SDR techniques. Similar to [274], the authors of [271] developed an AO algorithm based on FP and SDR techniques to address a weighted optimization of the SR and the TIP under the transmit power constraint. In addition, the authors of [269] examined the maximization of BPG under the SINR requirements for users and the target for both the ideal case with perfect CSI and known target location and the imperfect CSI case with uncertain target location. More recently, the authors in [280] also considered a SR maximization problem and jointly designed robust AN-aided BF and RIS phase shifts under the imperfect sensing parameters (distance and angle). Additionally, the authors of [281] employed multistatic cooperative sensing to determine the angle-of-arrival (AoA) and the localization of the eavesdropping target, thereby achieving previse beam alignment. They proposed a joint optimization of the BS-user association, the AN matrix, and the RIS phase shifts to maximize the sum SR, subject to the requirement of target sensing.

In order to overcome the performance limitations inherent to conventional fixed-position antennas (FPAs), the authors of [275] proposed the implementation of a movable antenna (MA) [291] and a RIS for a secure ISAC system. They attempted to maximize the sum rate of users by jointly optimizing transmit/receive BF, the reflection coefficients of RIS, and the positions of MAs at users, subject to a minimum communication rate requirement for users, a minimum radar sensing requirement, and a maximum information leakage to Eve. Furthermore, the authors of [276] considered RIS-UAV in secure ISAC and proposed a methodology to maximize the SR by jointly designing the UAV deployment, AN-aided transmit BF, and the RIS phase shifts, subject to the SNR requirement of the sensing target, the transmit power budget, and the constraints of RIS coefficients and UAV flight area. To address the non-convex problem, they developed an AO algorithm using SCA and SDR techniques.

Contrary to the aforementioned works, which consider a single sensing target, the authors of [277], [278] considered multiple sensing targets. In [277], they proposed a transmit BF and RIS phase shift design to maximize the SR under power constraints and a certain SNR requirement for each radar target. On the other hand, the authors in [278] sought to optimize the minimum weighted BPG by jointly designing the transmit BF, the AN matrix, and the RIS phase shifts. This optimization was performed subject to the communication QoS and security requirements, as well as the power budget. Furthermore, the authors in [279] considered a multi-cluttered environment and proposed a joint design of AN-aided transmit BF and RIS phase shifts by introducing the SCNR as a constraint of sensing performance. Unlike standard approaches, such as SDR, FP, and SCA, they proposed a Riemannian Manifold-based AO algorithm to address the problem.

*2) Active RIS-Assisted ISAC:* In general, a RIS with passive loads is designed to be passive or nearly passive, requiring no power when manipulating the EM signal impinging upon it with fixed reflecting coefficients. However, the passive nature of RIS imposes constraints due to double-fading attenuation, necessitating the utilization of cascaded channels composed

TABLE VII
SUMMARY OF PAPERS ON ACTIVE RIS-ASSISTED ISAC.

| Reference | User(s) | Eve(s) | Target(s) | KPIs (Com. / Sen.) | Eve's CSI | Opt. Variables | Optimization Techniques |
|-----------|---------|--------|-----------|--------------------|-----------|----------------|-------------------------|
| [292] | S | S | S | SR / SINR | Perfect | BF, RIS Phase Shifts | AO, MM, SDP, SCA |
| [293] | S | S | S | SR / SNR | Perfect | BF, Radar Receive Filter, RIS Phase Shifts | AO, MM, SDP |
| [294] | S | S | S (Eve) | SR / TIP | Perfect | BF, RIS Phase Shifts | AO, FOTE, MM, SDR |
| [295] | M | S | S (Eve) | SR / SNR | Perfect | BF, Radar Receive Filter, RIS Phase Shifts | AO, FP, MM |
| [296] | M | S | S (Eve) | SINR / BPG | Perfect | BF, RIS Phase Shifts | SCA |
| [297] | M | S | S (Eve) | SR / MI | Outdated | BF, Passive / Active RIS Phase Shifts, Spectrum Allocation | DRL (DQN and TD3 [298]) |
| [299] | M | M | M (Eve) | SR / ISLR | Perfect | Active RIS Phase Shifts, AN, BF, Power Splitting Ratio | AO, FOTE, SCA, SDR, Penalty Method |

S: Single, M: Multiple.

of transmitter-RIS and RIS-receiver subchannels [300]. To enhance the RIS-aided link to a satisfactory level, passive RIS demands a substantial number of reflecting elements, which requires a considerable surface area.

To address these issues, an active RIS has been proposed, in which each reflecting element is supported by a set of active-load impedances [301]. Compared with its passive counterpart, the active RIS is likely to be an active reflector, which directly reflects the incident signal with the power amplification in the EM level, and it still takes advantage of the RIS that no complex and power-hungry RF chain components are needed [302]. Recent studies have explored the application of active RIS in the context of PLS in ISAC, as outlined in Table VII.

*Single-User Scenario:* The employment of an active RIS within a secure ISAC system has been examined in the presence of a legitimate user, an eavesdropper, and a sensing target [292]. The authors formulated a SR maximization problem subject to power constraints at the BS and the active RIS, as well as radar SINR constraints, and optimized transmit BF and RIS reflection coefficients. The results of this study indicated that the active RIS demonstrated superiority over the passive RIS in terms of SR and sensing performance. In a related context, in [293], the SR maximization problem is formulated to jointly optimize transmit BF, the RIS phase shift matrix, and the radar received BF.

Meanwhile, existing works are based on the assumption that the signals from all links arrive at the destination concurrently. However, such an assumption does not hold in practice [303]. To address this challenge, the authors of [294] proposed delay alignment modulation (DAM) [304] for a secure active RIS-aided terahertz (THz) ISAC system that introduces a delay at the BS to ensure that multipath signals arrive at the user simultaneously without inter-symbol interference (ISI). To this end, they developed an AO algorithm by decomposing the original SR maximization problem into two subproblems, i.e., optimizations of transmit BF and active RIS phase shifts. They then applied the MM and SDR techniques to transform the non-convex subproblem into a convex one.

*Multi-User Scenario:* In [295], the authors investigated the design of an active RIS-assisted secure ISAC system in the presence of a malicious UAV as a sensing target. They addressed the SR maximization problem by jointly designing the transmit/receive BF and the reflection matrix of active RIS, subject to the constraint of the radar SNR threshold and total power budgets at the BS and the active RIS. The problem was decomposed into three sub-problems, and an iterative algorithm was developed based on FP and MM techniques.

The authors of [299] proposed secure BF for integrated sensing and simultaneous wireless information and power transfer (SWIPT) via active RIS. They formulated a joint optimization of the transmit BF, AN, power splitting ratios, and active RIS phase shifts to maximize the total harvested power while satisfying the constraints of sidelobe level ratio and secrecy rate. In order to address the aforementioned problem, the original problem was transformed into two sub-problems. In addition, two alternating optimization algorithms based on SDR and SCA were proposed.

In [297], the authors considered a hybrid active-passive RIS-enhanced ISAC system in vehicle-to-everything (V2X) networks, wherein V2V links share the spectrum resource occupied by vehicle-to-infrastructure (V2I) links. They optimized the sum SR of V2I links by jointly designing the transmit BF of roadside units, the radio spectrum reuse scheme of V2X links, and the active/passive reflection coefficients of hybrid RIS. To address the non-convex problem and the dynamic system, they developed a hierarchical twin delayed DDPG method to learn the secure BF and spectrum sharing strategies.

Recently, the authors of [296] presented a comparative analysis of passive and active RISs in secure ISAC systems. Specifically, they jointly optimized transmit BF and RIS coefficients to maximize the BPG under the SINR constraints of the communication users and the eavesdroppers, as well as the power budget. The results of this study demonstrated that active RIS systems exhibit a substantial enhancement in terms of BPG when compared to their passive counterparts. However, this enhancement is accompanied by an increase in hardware costs and computational complexity.

*3) STAR-RIS-Assisted ISAC:* In the context of passive RISs, the location of both the transmitter and receiver is confined to the same side of the RIS [312]. However, this geographical restriction is not always feasible in practice, impeding the flexibility and effectiveness of RISs. To address this limitation, a novel concept, termed simultaneously transmitting and reflect-

TABLE VIII
SUMMARY OF PAPERS ON STAR-RIS-ASSISTED ISAC.

| Reference | User(s) | Eve(s) | Target(s) | KPIs (Com. / Sen.) | Eve's CSI | Opt. Variables | Optimization Techniques |
|-----------|---------|--------|-----------|--------------------|-----------|----------------|-------------------------|
| [305] | M | S | S | SR / SNR | Perfect | BF, Radar Waveform, STAR-RIS Phase Shifts | AO, FOTE, SCA, SROCR |
| [306] | M | S | S (Eve) | SR / SNR | Perfect | BF, Radar Receive Filter, STAR-RIS Phase Shifts | DRL (DDPG and SAC) |
| [307] | M | S | S (Eve) | SINR / BPG | Perfect | BF, IOS Phase Shifts | AO, SDR |
| [308] | M | S | S (Eve) | COP / SOP | Perfect | Time Allocation | - |
| [309] | M | M | M (Eve) | User's Rate / BPM SINR / BPG | Imperfect | BF, AN, STAR-RIS Phase Shifts | AO, BCD, FOTE, Penalty Method, SCA, S-procedure |
| [310] | M | M | M (Eve) | SR / SNR | Perfect | BF, STAR-RIS Phase Shifts | DRL (DDPG and SAC) |
| [311] | M | S | S (Eve) | SR / CRB | Perfect | BF, Sensing Waveform, STAR-RIS Phase Shifts, | ALM, BCD, SDR |

S: Single, M: Multiple, COP: Communication Outage Probability, SOP: Sensing Outage Probability.

ing (STAR)-RISs, was proposed in [313], [314]. In particular, the wireless signal incident on an element of a STAR-RIS from either side of the surface is divided into two parts [315]. One part (reflected signal) is reflected to the same space as the incident signal, i.e., the reflection space, and the other part (transmitted signal) is transmitted to the opposite space as the incident signal, i.e., the transmission space. As demonstrated in [313], the transmission and reflected signals of a STAR-RIS element can be reconfigured through the manipulation of the electric and magnetic currents. This reconfiguration is enabled by two generally independent coefficients, termed the transmission and reflection coefficients, respectively.

STAR-RIS employs three distinct protocols: energy-splitting (ES), mode switching (MS), and time switching (TS) [314]. These protocols empower STAR-RIS to transmit, reflect, or both of these functions concurrently. In the ES and MS protocols, the incident signal is bifurcated into two portions by the STAR-RIS. In contrast, a TS-based STAR-RIS utilizes a switch to allocate discrete time periods for its transmissions and reflections.

An illustration of a STAR-RIS-aided ISAC system is presented in Fig. 9, where a communication user and a target are situated in the transmission and reflection spaces, respectively. A recent study has examined the application of STAR-RIS in the context of PLS problems, the details of which are outlined in Table VIII.

In the papers [305]–[308], [311], the authors considered a STAR-RIS-assisted ISAC system with a single sensing target. The authors of [305] considered the ES protocol and the joint design of transmit BF at the BS and transmission/reflection BF at the STAR-RIS to maximize the SR. This was subject to the sensing SNR constraint, the transmit power constraint, and the STAR-RIS's energy constraint. The problem was addressed using the SCA and sequential rank-one constraint relaxation methods. In [306], a similar SR maximization problem was addressed by using SAC and DDPG approaches.

On the other hand, the authors of [307] proposed a strategy to maximize the BPG while satisfying the requirements of minimum SINR for users and maximum SINR for the malicious target. This was achieved by jointly optimizing transmit BF and the phase shift matrices of STAR-RIS (referred to as intelligent omni-surface (IOS) in [307]). Furthermore, the authors of [311] proposed a joint design of transmit BF and RIS phase shift coefficients to minimize the CRB under the constraints of a SR and power budget.

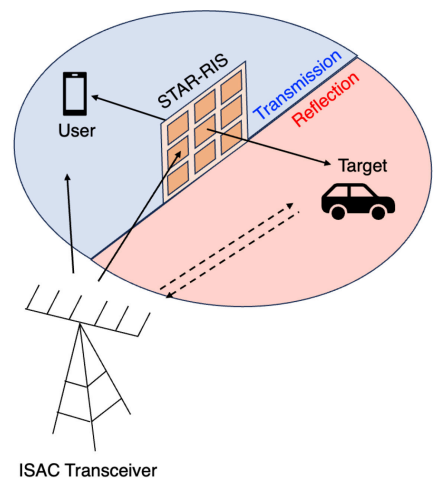

Fig. 9. A STAR-RIS-assisted ISAC network.

In contrast to diagonal RISs, which feature connections between each RIS element and a single load disconnected from the others, beyond diagonal RIS has been proposed to allow for inter-element connections. Specifically, the fully-connected (FC)-RIS represents a notable case that offers the maximum BF gain [316]. In [308], the authors explored the STAR-FC-RIS, which integrates FC-RIS with the TS-STAR-RIS. They derived closed-form expressions for the sensing outage probability, communications outage probability, and communications intercept probability. They then exploited an average of the three probabilities to obtain an optimized time allocation of the TS-based STAR-FC-RIS.

Contrary to the aforementioned works, the papers [309], [310] considered multiple sensing targets (eavesdroppers). In [310], the authors considered a joint optimization of the transmit BF, STAR-RIS transmission/reflection coefficients, and receive filters, with the objective of maximizing the long-term average SR for communication users. Constraints were imposed to guarantee minimum echo SNRs for sensing and to align with the achievable rate requirements of communi-cation users. To address the non-convex problem, two DRL algorithms, namely, DDPG and SAC, are proposed.

In [309], the authors investigated two archetypal ISAC scenarios: communication enhancement and communication-sensing fusion scenarios. They formulated the AN-aided robust and secure BF problems for these scenarios, considering imperfect CSI. Additionally, they examined the BF design for several STAR-RIS protocols, including the ideal ES, the coupled phase-shift ES, and the MS protocols.

*Summary:* A review of the existing literature reveals that the utilization of RIS has been shown to offer significant advantages in terms of communication security and sensing performance when compared to conventional ISAC systems without RIS. Moreover, numerous studies have demonstrated that active RIS and STAR-RIS have the potential to surpass passive RIS in terms of PLS and sensing performance, albeit at the expense of increased hardware cost. However, existing studies presuppose a specific RIS deployment, such as the locations (potentially mounted on UAVs) and the number of reflecting elements. In practical heterogeneous networks, these factors will be meticulously designed. Furthermore, large RISs operating at high carrier frequencies possess an expanded near-field region. Consequently, near-field communication and sensing will become more prevalent in RIS-assisted ISAC systems [317], which complicates channel estimation and BF designs.

### E. NOMA/RSMA-Aided ISAC

Despite the exponentially increasing number of users in next-generation wireless networks, orthogonal transmission schemes are only capable of supporting a limited number of users for a given set of orthogonal resources. This limitation necessitates the development of non-orthogonal transmission schemes for next-generation multiple access.

*1) NOMA-ISAC:* The power domain (PD)-NOMA has emerged as a promising candidate for next-generation multiple access schemes [330]. The fundamental concept of PD-NOMA revolves around the utilization of shared time, frequency, and code resources to serve multiple users while distinguishing them based on power domain parameters [331], [332]. The two key technologies in PD-NOMA are superposition coding and successive interference cancellation (SIC), which have been proven to be capacity-achieving in the single-antenna BS and multiple-access channel (MAC). Recently, the idea of NOMA has been studied in the context of ISAC [333]–[336]. In addition, a number of papers have investigated PLS problems in NOMA-ISAC systems. A summary of these studies is provided in Table IX.

In [318], [337], the authors investigated a secure NOMA-ISAC system in the presence of a single target that acts as an eavesdropper. They considered a joint transmit BF design to maximize the sum SR for multiple users via AN while satisfying the demodulation order, the SINR constraint for NOMA, user's rate thresholds, radar SNR, and total power budget. The authors of [328] considered a single target and eavesdropper in the near-field scenario. They proposed a joint transmit BF design to maximize the sum SR, subject to constraints of the CRB, rates of communication users, and transmit power. In contrast, the authors of [319] considered a scenario involving separate sensing targets and eavesdroppers, proposing a transmit BF design to maximize the user's sum rate under the constraints of the decoding order and radar performance. Furthermore, the authors of [323] considered multiple targets (eavesdroppers) in a coordinated multi-point transmission (CoMP)-NOMA-based ISAC system, where multiple BSs are coordinated to enhance PLS and sensing performance. They formulated the joint radar association and transmit BF design to maximize the minimum BPG, subject to maximum power constraint and security requirements.

Contrary to the aforementioned works, the authors in [322], [324] took into account imperfect CSI of the eavesdropper. In [322], the authors considered a multi-carrier (MC)-NOMA-based ISAC system in the presence of an active eavesdropper with imperfect CSI. They optimized the subcarrier allocation and AN-aided transmit BF to maximize the minimum user's rate while ensuring requirements for information leakage to the eavesdropper and sensing performance in terms of CRB. In addition, in [324], the authors considered the scenario of imperfect CSI of the eavesdropper in a LEO satellite NOMA-ISAC system, and they optimized transmit BF to maximize the sum SR while considering constraints on user's QoS, the NOMA decoding order, radar SNR, and total power budget. The authors of [329] also addressed the imperfect CSI of the eavesdropper in a two-user cooperative NOMA with SWIPT. Specifically, they proposed a two-phase secure transmission mechanism in an energy-limited scenario, where the cell-center user employs SWIPT to harvest energy used for forwarding the cell-edge user's information. They formulated a joint optimization of BF vectors, AN covariance matrix, power splitting factor for the first and second phases, and power allocations for communication and AN signals is formulated, with the objective of minimizing the transmit power of the BS.

Furthermore, the authors of [320], [321] investigated the scenario involving both imperfect CSI and SIC. In [321], the authors proposed secure BF and power allocation methodologies based on user clustering under imperfect CSI and SIC. The problem was formulated as a TDP maximization problem under the constraints of the SR, the decoding order, and the power budget. Similarly, the authors in [320] proposed a robust BF design under imperfect CSI and target angles to maximize the energy efficiency under the security and sensing performance constraints in a NOMA-ISAC system with two users, namely, communication-centric and radar-centric users [338].

Meanwhile, the papers [325]–[327] independently studied secure RIS-assisted NOMA-ISAC systems at nearly the same time. The authors of [326], [327], [339] explored the joint optimization of transmit BF and passive RIS phase shifts, with the objective of maximizing the sum SR. This was achieved under the constraints of the NOMA decoding order, the user QoS, the transmit power, and the radar BPG or SNR. To address this problem, they both proposed an AO algorithm leveraging the first-order Taylor expansion, the SOC techniques [121], and SCA. Furthermore, in [325], [340], the authors explored

TABLE IX
SUMMARY OF PAPERS ON PLS FOR NOMA-ISAC.

| Reference | User(s) | Eve(s) | Target(s) | KPIs (Com. / Sen.) | Eve's CSI | Opt. Variables | Optimization Techniques |
|---|---|---|---|---|---|---|---|
| [318] | M | S | S (Eve) | SR / SNR | Perfect | BF, Jamming | FOTE, SCA, SOC |
| [319] | M | S | S | User's Rate / BPM | Unkown | BF | BCD, FOTE |
| [320] | D | S | S (Eve) | SR / SINR | Imperfect | BF | Penalty Method, SDR, S-Procedure |
| [321] | M | M | M (Eve) | SR / TDP | Imperfect | BF, Radar Waveform, Power Allocation | AO, Iterative RM [211], SDR, S-Procedure |
| [322] | M | S | S | User's Rate / CRB | Imperfect | BF, AN, Subcarrier Allocation | Big-M Formulation, FP, LD, Penalty Method, SCA, SDR |
| [323] | M | M | M (Eve) | SR / BPG | Perfect | BF, AN, Radar Association | Accelerated Stochastic CD, FOTE, Penalty Method |
| [324] | M | S | S (Eve) | SR / SNR | Perfect / Imperfect | BF, AN | AM-GM Inequality, FP, FOTE, Penalty Method, SCA, SDR, S-Procedure |
| [325] | M | S | S (Eve) | SR / BPG | Perfect | BF, Jamming, STAR-RIS Phase Shifts | AO, FOTE, SCA, SOC |
| [326] | M | S | S (Eve) | SR / BPG | Perfect | BF, RIS Phase Shifts | AO, FOTE |
| [327] | M | S | S (Eve) | SR / SNR | Perfect | BF, Jamming, RIS Phase Shifts | AO, FOTE, SOC |
| [328] | M | S | S | SR / CRB | Perfect | BF, Radar Waveform | FOTE, SCA, SDR |
| [329] | D | S | S (Eve) | SR / SINR | Imperfect | BF, AN, Power Splitting Factor, Power Allocation | AO, SDR, S-Procedure |

S: Single, M: Multiple.

STAR-RIS-assisted secure NOMA-ISAC systems, where the sensing target and users are located in the reflection and transmission regions of STAR-RIS, respectively. They also considered the maximization of the sum SR through the joint optimization of the transmit BF and the transmission/reflection coefficients of STAR-RIS, while ensuring the minimum BPG requirement. Similar to [325]–[327], the problem was addressed by an AO algorithm employing the first-order Taylor expansion, the SOC constraints, as well as SCA techniques.

*2) RSMA-ISAC:* Rate-splitting multiple access (RSMA) is a recently developed multi-user scheme relying on the rate-splitting principle [349], [350]. In this scheme, the transmitter divides part of each user's message (i.e., the private message) into a common message intended for all served users. The private messages remaining after this step, along with the constructed common message, are then transmitted via BF. When decoding the common message, the receivers treat the private messages of all users as interference and subtract the common message from the received signal (i.e., SIC). The intended private message is then decoded and combined with part of the decoded common message. Recent advancements in the integration of RSMA and ISAC have been examined in [351]. A summary of recent studies that address PLS problems in RSMA-ISAC systems is provided in Table X.

In [341], the authors examined RSMA within the context of an ISAC system, where the common messages act as interference for eavesdroppers, akin to AN. They proposed a joint design of transmit BF and a common rate of RSMA as a weighted optimization to achieve trade-offs between communication and sensing performances. To address the non-convex nature of the problem, they developed an SCA with a penalty algorithm that employs an iterative approach, performing convex approximations and introducing penalties. Recently, the authors of [342] investigated a cognitive radio network scenario involving the coexistence of ISAC and multicast communications based on RSMA. In this scenario, they proposed a transmit BF design aimed at maximizing the security energy efficiency, defined as the ratio of SR to power consumption, by leveraging green interference from multicast communication signals. Additionally, the authors of [346] proposed a secure communication method for the RSMA-ISAC system that utilizes fluid antennas. They formulated a joint design of transmit BF and antenna positions to maximize the secrecy rate under both perfect and imperfect CSI conditions. They developed AO-based solutions to the non-convex optimization problem, where the original problem was decomposed into optimizations of BF and fluid antenna positions.

Furthermore, the papers [343], [344], [347], [348] considered secure RIS-assisted RSMA-ISAC systems. The authors in [343], [344] assumed a single sensing target, which was considered a potential eavesdropper in [343]. Specifically, the authors in [343] considered the bounded CSI error model for the eavesdropper and proposed a strategy to maximize the worst-case minimum SR by jointly designing transmit BF and the passive RIS phase shifts. On the other hand, the authors in [344] assumed that the eavesdropper appears in a specific region with an unknown position and that AN is employed to enhance security. They formulated the joint optimization of the common rate allocation of RSMA, the AN vector, transmit BF, active RIS reflection coefficients, and radar receive filter to maximize the minimum privacy SR, while ensuring the minimum requirement of common SR, the minimum radar output SNR requirement, and power budget constraints of active RIS and BS.

On the other hand, the authors in [347], [348] considered a scenario involving multiple sensing targets that act as eavesdroppers. In [347], the authors considered an active RIS comprising reflecting elements and dedicated sensors within a context where the BS possesses imperfect CSIs of the

TABLE X
Summary of papers on PLS for RSMA-ISAC.

| Reference | User(s) | Eve(s) | Target(s) | KPIs (Com. / Sen.) | Eve's CSI | Opt. Variables | Optimization Techniques |
|---|---|---|---|---|---|---|---|
| [341] | M | S | M | SR / BPG | Imperfect | BF, Common Rate of RSMA | FOTE, Penalty Method, SCA, S-Procedure |
| [342] | M | S | | SEE / SCNR | Imperfect | BF, Radar Receiver Filter | AO, FOTE, MM, SCA, SDP |
| [343] | M | S | S (Eve) | SR / BPG | Imperfect | BF, RIS Phase Shifts, Common Rate of RSMA | FOTE, Penalty Method, SCA, SDR, S-Procedure, |
| [344] | M | S | | SR / SNR | Imperfect | BF, AN, Radar Receive Filter, Active RIS Phase Shifts, Common Rate of RSMA | AO, Lemmas of Matrix Inversion and Matrix Fractional Functions [345], MM, SCA, SOTE |
| [346] | M | S | S (Eve) | SR / BPG | Perfect / Imperfect | BF, Antenna Position | AO, SCA, SOTE, S-Procedure |
| [347] | M | M | M (Eve) | SOP / SINR | Imperfect | BF, Radar Waveform, RIS Phase Shifts | AO, BTI, SCA, SDR, SOTE |
| [348] | M | M | M (Eve) | SR / TDP SOP / BPM | Imperfect | BF, Common Rate of RSMA, Timeslot Duration | AO, BCD, BTI, SCA, SOC, S-procedure |
| [348] | M | M | M (Eve) | SR / TDP SOP / BPM | Imperfect | BF, Common Rate of RSMA, Timeslot Duration | AO, BCD, BTI, SCA, SOC, S-procedure |

S: Single, M: Multiple, SEE: Security Energy Efficiency.

eavesdroppers. They formulated the radar SINR maximization problem under the SOP constraints for the users. In [348], the authors considered a time-division-based ISAC system employing a transmissive RIS-empowered BS [352] and exploited the common stream of RSMA as both useful signals and AN. They jointly optimized the common/private stream BF and time slot duration variable to maximize the sum SR under the channel uncertainties for both users and targets.

*Summary:* It has been demonstrated that, in comparison with orthogonal multiple access, NOMA can achieve higher communication security while ensuring the sensing performance. Furthermore, the use of RIS in NOMA/RSMA-ISAC has been shown to provide additional optimization DoF as well as additional virtual LOS links, thereby potentially augmenting both PLS and sensing capabilities. However, the joint optimization of transmit BF, RIS phase shifts, and NOMA/RSMA parameters is generally very complex. Additionally, passive RIS may hinder perfect CSI acquisition, whose impact should be considered in the design of NOMA/RSMA systems. Furthermore, the existing literature on this subject is limited in scope: it considers only a downlink scenario and external eavesdropper(s). Consequently, future research directions may encompass the following: uplink ISAC [334], cooperative NOMA [353] in ISAC [354], and security design against internal eavesdropper(s).

### F. Symbol-Level Precoding

The directional modulation technique, which is based on the principle of constructive interference (CI) and destructive interference (DI) [363]–[366], has been recently employed to design the transmit signal at a symbol level in secure ISAC systems [355], [357], [360]. A summary of these works can be found in Table XI.

The directional modulation technique enhances the SR of ISAC systems by leveraging the CI and DI techniques, which enable the direct alteration of signals' amplitude and phase at both users and eavesdroppers. For the sake of clarity, the constructive regions of QPSK and 8-PSK are illustrated as blue regions in Fig. 10. Given the knowledge of the channel and all user's data readily available at the transmitter, CI and DI move the received symbols into the CI regions and away from the CI regions, respectively. As a result, this paradigm promotes an enhanced SER for users while deteriorating the decoding probability at eavesdroppers.

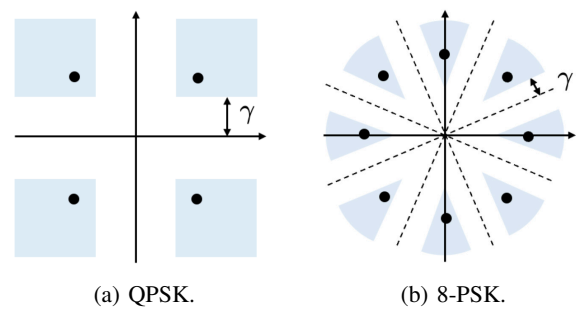

(a) QPSK.　　　　(b) 8-PSK.

Fig. 10. Constructive regions (blue areas) for QPSK and 8-PSK. The scalar $\gamma$ denotes the threshold distance to the decision variables, that relates to the SNR constraint [363].

The authors [355] proposed a BF design for phase-shift keying (PSK) modulation that exploits CI to ensure the QoS of legitimate users and DI to disrupt the communication link with an eavesdropper. They formulated the BF design as CRB minimization under the CI and DI constraints, as well as the transmit power budget.

In contrast, the authors [357], [367] considered the radar SINR maximization problem under the constraints of CI and DI. Moreover, the authors in [360] considered the estimation task of random parameters of multiple sensing targets, where the prior distribution of parameters is presumed to be known *a priori*. To this end, they evaluated and attempted to minimize BCRB while guaranteeing a predefined SNR at users by employing CI and DI techniques. The results demonstrated that the proposed symbol-level precoding exhibits superior performance in terms of the performance trade-off between communication security and sensing when compared with the

| Reference | User(s) | Eve(s) | Target(s) | KPIs (Com. / Sen.) | Eve's CSI | Opt. Variables | Optimization Techniques |
|-----------|---------|--------|-----------|--------------------|-----------|-----------------|-------------------------|
| [355] | M | S | S (Eve) | CI-DI Const. / CRB | Perfect | BF | Armijo Line Search [356], FOTE, SCA |
| [357] | M | M | M (Eve) | CI-DI Const. / SINR | Perfect / Imperfect | Transmit Waveform, Radar Receive Filter | Big-M Formulation, FOTE, FP, SCA, SDR, SOA [81] |
| [358] | M | M | M (Eve) | CI-DI Const. / BPM, BPG, PAPR | Perfect | Transmit Waveform, STAR-RIS Phase Shifts | AO, Courant's Penalty Method [359], DM, MM |
| [360] | M | M | M (Eve) | SNR / BCRB | Statistical | Transmit Waveform | FOTE, Line Search, SCA |
| [361] | M | S | S (Eve) | User's SNR / CRB | Perfect | Transmit Waveform, RIS Phase Shifts | AO, FOTE |
| [362] | M | M | M (Eve) | CI-DI Const. / SINR | Perfect | Transmit Waveform, Radar Receive Filter, RIS Phase Shifts | ADMM, BCD, FOTE, Penalty Method, SCA |

S: Single, M: Multiple.

conventional block-level counterpart.

In a related study, the authors of [361] proposed a secure symbol-level precoding method for an RIS-aided ISAC system in the presence of a single sensing target. This problem was formulated as a weighted optimization problem, with the objective being to maximize the user's rate while minimizing the CRB under the constraints of CI and DI that ensure communication QoS for communication users and security for the eavesdropper. In addition, the authors of [358], [368] proposed symbol-level precoding for a STAR-RIS-assisted ISAC system. They formulated the joint design of transmit waveform and transmission/reflection coefficients of STAR-RIS to maximize the BPG under the constraints of the similarity between the desired and the designed beampatterns, PAPR, and CI and DI constraints. Furthermore, the authors of [362] considered RIS with finite phase shifts and multiple sensing targets. In particular, they developed a methodology for jointly designing a transmit waveform, a radar receive filter, and discrete phase shifts of the RIS, with the objective of maximizing the minimum radar SINR. This was achieved under the constraints of CI and DI, the power budget, and the required PAPR level. To address the problem, the authors developed a BCD algorithm, incorporating SCA, a penalty technique, and the ADMM algorithm [186].

*Summary:* The preceding works have demonstrated that symbol-level CI and DI techniques enhance communication security and sensing performance of conventional block-level precoding by leveraging the known multi-user interference (MUI) as a source of useful signal. However, there are some challenges in the optimization of precoding designs. Firstly, all the aforementioned works consider PSK signaling. While the concept of CI and DI can be extended to general constellations in principle, such as quadrature amplitude modulation (QAM), it necessitates a meticulous design of optimization constraints. Furthermore, while perfect CSI of communication users is assumed in the aforementioned works, the optimal robust precoding design remains a challenging problem to be investigated.

### G. Mobile Eavesdropper

While the majority of existing works assume that the locations of eavesdroppers are fixed, some papers have considered the case of mobile eavesdroppers [114], [115], [369], [370]. A summary of these studies can be found in Table XII.

In [114], the authors considered a secure ISAC system with a mobile aerial eavesdropper. They adopted an EKF to predict the eavesdropper's motion state. Utilizing tracking information, they formulated an optimization problem to jointly design the radar signal and receive BF, with the objective of maximizing the minimum SINR of users under the constraints of eavesdropper's SINR, tracking MSE, and power budget. The problem was solved through the application of an AO-based algorithm, which iteratively employed FP and SDP. The authors of [115] also adopted the EKF to predict the mobile eavesdropper's CSI in a near-field ISAC system. They formulated a multi-objective optimization to simultaneously optimize power consumption, the number of securely served users, and tracking performance, while guaranteeing the achievable rates of users and the eavesdropper. They proposed a globally optimal solution based on GBD, as well as a low-complexity solution based on ZF BF and SCA.

In [369], the authors considered an ISAC system in which a BS transmits both public and confidential messages to UAVs with disparate security levels, concurrently executing target tracking. Specifically, public messages are transmitted to all the UAVs, while confidential messages are directed to the UAVs with a higher security level. Additionally, AN is exploited to minimize the SINR of the confidential messages received by the UAV with a lower security level. In this setup, the authors addressed the BPG problem under the constraints of the public and confidential message rates for time-variant and time-invariant channels.

In [370], the authors considered a mobile proactive adversarial target, designated as a malicious UAV. They formulated the interaction between the legitimate network and the malicious UAV as a non-cooperative Stackelberg game [373], wherein the malicious UAV strategically adjusts its trajectory to eavesdrop, while the legitimate network dynamically allocates resources to ensure requisite SRs and sensing performance. The problem is addressed by employing a SCA-based solution for

TABLE XII
SUMMARY OF PAPERS ON MOBILE EAVESDROPPERS IN ISAC.

| Reference | User(s) | Eve(s) | Target(s) | KPIs (Com. / Sen.) | Eve's CSI | Opt. Variables | Optimization Techniques |
|---|---|---|---|---|---|---|---|
| [114] | M | S | S (Eve) | SINR / Tracking MSE | Unknown | Radar Waveform, Radar Receive Filter | AO, FP, SDR |
| [115] | M | S | S (Eve) | SR / Tracking MSE | Unknown | BF, Radar Waveform, User Scheduling | GBD, SCA, Schur Complement, SDR, S-Procedure, ZF |
| [369] | M | M | M (Eve) | SR / BPM | Perfect / Imperfect | BF, AN | FOTE, SCA, SDP |
| [370] | M | S | S | SR / CRB | Unkown Range | Transmit / Receive BF, AN, Power Allocation, UAV Trajectory | FOTE, DRL (Double DQN), SCA, SDR |
| [371] | M | S | S (Eve) | User's Rate / CRB | Unknown | Transmit Waveform, RIS Phase Shifts | AO, FP, MO, SCA |
| [372] | M | M | M (Eve) | SR / SINR | Unknown | BF | FOTE, Deep Learning (LSTM), SCA, ZF |

S: Single, M: Multiple.

network resource optimization and DRL (double DQN) for UAV trajectory planning.

In contrast to the aforementioned works, the authors of [372] considered multiple mobile targets (eavesdroppers). Instead of employing the EKF, they utilized a long short-term memory (LSTM) model to predict eavesdropper's channels. Additionally, they adopted meta-learning to provide generalization capability to any trajectory with a few fine-tuning steps. They developed two secure BF algorithms to maximize the sum SR based on SCA and ZF techniques, using the predicted eavesdropper's channels.

Furthermore, the authors of [371] considered secure transmission for an RIS-based backscatter system, wherein an ISAC BS simultaneously functions as a carrier emitter to active internet of things (IoT) devices and a sensing node for tracking an aerial eavesdropper. Leveraging the predicted CSI of the eavesdropper by sensing, they formulated a sum-rate maximization problem. To address this problem, they proposed a joint optimization of transmit waveform and RIS phase shifts, subject to the constraints of the eavesdropper's SINR and CRB in eavesdropper's angle estimation. They developed an FP-based AO algorithm, where SCA and MO techniques are employed to tackle the non-convex constraints.

*Summary:* In scenarios involving a mobile eavesdropper, the EKF technique has been extensively adopted for predicting and acquiring the eavesdropper's CSI. Furthermore, the tracking performance of the EKF may be improved by deep learning techniques, such as LSTM [374]. The tracking information can be utilized for secure transmission design, such as BF, to enhance PLS. However, existing works assume the existence of a LOS link between the BS and the mobile eavesdropper, overlooking the potential for non-line-of-sight (NLOS) paths. This may be addressed by deploying a RIS to establish virtual LOS pathways. Furthermore, exploring general cases involving multiple mobile (potentially active) eavesdroppers is a promising avenue for future research.

### H. Jamming and Anti-Jamming Strategies

Several papers have examined jamming and anti-jamming strategies for secure ISAC systems. A summary of these works can be found in Table XIII. Typical jamming and anti-jamming

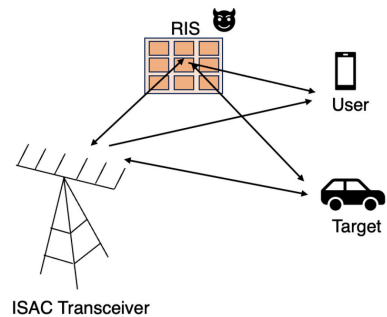

(a) A jamming scheme with RIS.

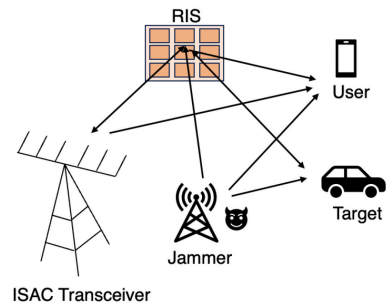

(b) An anti-jamming scheme with RIS.

Fig. 11. Jamming and anti-jamming strategies with RIS.

scenarios for an ISAC system based on RIS are illustrated in Fig. 11.

In [376], the authors considered a secure ISAC system in the presence of both an eavesdropper and a jammer who emits a jamming signal to disrupt both the sensing and communication of ISAC signals. In this context, they sought to maximize the SR by optimizing transmit BF and radar receive filter while satisfying the radar SINR requirements and the transmit power budget. To address this problem, an AO-based algorithm was developed based on SDR and BCD methods.

As in Fig. 11b, the authors [377], [382] proposed an anti-jamming scheme for an ISAC system via aerial RIS. Their objective was to maximize the achievable sum rate of communication users by jointly optimizing the transmitting

TABLE XIII
Summary of papers on jamming and anti-jamming strategies.

| Reference | User(s) | Eve(s) | Target(s) | KPIs (Com. / Sen.) | Eve's CSI | Opt. Variables | Optimization Techniques |
|-----------|---------|--------|-----------|--------------------|-----------|----------------|-------------------------|
| [375] | S | S | S (Eve) | User's Rate / ADPAR | - | Jamming Waveform | PGD |
| [376] | M | S | S | SR / SINR | Perfect | BF, Radar Waveform, Radar Receive Filter | BCD, FOTE, Penalty Method, SDR |
| [377] | M | - | S | User's Rate / SINR | Perfect | BF, RIS Phase Shifts, Aerial RIS Deployment | AO, FOTE, Penalty Method, SCA, SDR |
| [378] | M | - | S | User's Rate / CRB | Perfect | BF, STAR-RIS Phase Shifts | AO, FOTE, FP, SCA, SDR |
| [379] | M | - | S | SINR / BPM | - | Transmit Waveform | SDR |
| [380] | M | - | M | SR / SINR | - | BF, RIS Phase Shifts | MO, SDR |
| [381] | M | S | S (Eve) | SR / BPM | Perfect | Transmit / Jamming Waveform, RIS Phase Shifts | AO, FOTE, FP, LD, MM |

S: Single, M: Multiple.

BF at the ISAC BS, as well as the phase shift matrix and deployment of aerial RISs, while satisfying the radar echo SINR requirement for target sensing. On the other hand, the authors in [378] proposed an anti-jamming scheme via STAR-RIS in the presence of a jammer, which is a sensing target. Specifically, they jointly optimized transmit BF and the transmission/reflection coefficient matrices of STAR-RIS to maximize the achievable rate under the constraints of power budget and CRB.

In [379], the authors investigated the impact of DISCO RIS-based fully-passive jamming (FPJ) on an ISAC system, as depicted in Fig. 11a. FPJ is capable of launching jamming attacks without relying on either jamming power or CSI. The concept of FPJ was first proposed in [383], where an adversarial RIS with random and time-varying reflective coefficients acts like a "DISCO ball" and is therefore called a DISCO RIS [384]. Specifically, they formulated the ISAC waveform design problem as a weighted optimization to achieve a trade-off between communication and sensing performances. Furthermore, the authors in [380] took the perspective of an attacker and optimized a RIS-based active attack against a DFRC system. Specifically, the attacker equipped with a malicious RIS optimized the RIS phase shifts to reduce the radar echo SINR and SR. In addition, the authors of [375] proposed a novel sensing-resistant jamming strategy from the attacker's perspective. This strategy aims to achieve maximum disruption to transmission while simultaneously deceiving the receiver's detection of the attacker's direction. The authors introduced angular-domain peak-to-average ratio (ADPAR) at the receiver as a sensing metric for the direction estimation and designed jamming for minimizing Bob's achievable rate under ADPAR constraints.

Recently, the authors of [381] examined two eavesdropper deployment scenarios: the first was the deployment of eavesdropper within the main lobe, and the second was the deployment of eavesdropper within the side lobes. In the former scenario, the sensing target was considered an eavesdropper, and cooperative jamming was introduced to disrupt the eavesdropper. They then optimized the transmit and jamming waveform to maximize the secrecy rate. In the latter scenario, they considered the case of a separate eavesdropper located between the ISAC BS and communication users, and employed RIS to enhance PLS performance. In this case, they developed a joint optimization of transmit waveform and RIS phase shift to maximize the secrecy rate.

*Summary:* The efficacy of RIS in both jamming and anti-jamming strategies for ISAC has been demonstrated. However, several ideal assumptions have been made in the existing literature. For instance, in [380], it is assumed that the attacker has established synchronization with the DFRC system and acquired statistical CSI of legitimate channels. Additionally, the impact of imperfect CSI was ignored in the aforementioned literature. Consequently, expanding the current research to encompass practical scenarios, characterized by imperfect synchronization and CSI, is identified as a crucial future direction. In addition, a comprehensive investigation into the various types of jamming attacks and anti-jamming techniques is suggested [385], [386].

### I. UAV-ISAC

The provision of flexible coverage for a substantial number of ground devices has led to significant interest in UAV BS. In comparison to the conventional ground BSs, UAV BSs possess the capability to be deployed in the three-dimensional space, thereby enhancing the ISAC performance [395]–[398]. However, communication security remains a challenging aspect for UAV-ISAC communications [387], [388], [399]. As a result, numerous papers have addressed the security issue, which are outlined in Table XIV.

*1) Single-User Scenario:* In [387], the authors considered a UAV-ISAC system that transmits messages to an IoT device on the ground while performing sensing to detect a potential eavesdropper. They optimized the UAV trajectory and transmit BF to maximize the SR under the sensing SNR constraint via the SCA algorithm.

In the meantime, the authors of [388] consider a dual-functional eavesdropper that attempts to intercept the signals intended for both sensing and communication. They sought to optimize the average SR for the communication user by jointly designing the UAV trajectory and the transmit BF, while ensuring sensing performances for both the target and the eavesdropper. To tackle the non-convex optimization problem, they proposed an AO in conjunction with SCA and SDR methods.

TABLE XIV
SUMMARY OF PAPERS UAV-ISAC.

| Reference | User(s) | Eve(s) | Target(s) | KPIs (Com. / Sen.) | Eve's CSI | Opt. Variables | Optimization Techniques |
|---|---|---|---|---|---|---|---|
| [387] | S | S | S (Eve) | SR / SNR | Perfect | BF, Radar Receive Filter, UAV Trajectory | FOTE, MM, SCA |
| [388] | S | S | D | SR / BPG | Perfect | BF, Radar Waveform, UAV Trajectory | AO, SCA, SDR |
| [389] | S | S | S | SR / SNR | Imperfect | BF, Radar, Receive Filter, RIS Phase Shifts, UAV Trajectory | BCD, FOTE, MM, SCA, SDR |
| [390] | M | S | S (Eve) | # of SSUs / Tracking MSE | Unknown | Resource Allocation, BF, AN, UAV Online Navigation | SDR |
| [391] | M | M | M (Eve) | SR / SNR | Perfect | User Scheduling, Transmit Power, UAV Trajectory | AO, FOTE, SCA |
| [392] | M | M | M (Eve) | SINR / BPG | Perfect | BF, UAV Position | AO, FOTE, FP, SCA, SDR |
| [393] | M | S | M | User's Rate / TIP | Unknown | Power Allocation, User / Target Scheduling, RIS Phase Shifts, UAV Trajectory and Velocity | AO, BCD, FP, RCG, SCA |
| [394] | M | S | D | User's Rate / SINR | Perfect | BF, Radar Receive Filter, Jamming UAV Trajectory | AO, FOTE, SCA, SDR |

S: Single, M: Multiple, SSUs: Securely Served Users.

A significant constraint of the UAV-ISAC system pertains to the potential obstruction of the LOS communication links with the ground users/targets. To address this limitation, a passive RIS-aided UAV-ISAC system was proposed in [389]. The authors attempted to optimize the SR under the constraints of power budget and sensing SNR by jointly designing the UAV trajectory, RIS phase shifts, and transmit/receive BF. To this end, they developed the BCD algorithm based on SCA, SDR, and MM techniques to address the original non-convex problem.

*2) Multi-User Scenario:* In [390], [400], the authors considered a scenario in which an ISAC-UAV transmits confidential information to multiple ground users while simultaneously jamming and tracking an eavesdropping UAV via EKF. Leveraging the sensing information, user scheduling and transmit precoding are optimized to maximize the number of securely served users while satisfying the tracking MSE constraint and the QoS requirements of legitimate users and UAV eavesdropper.

In contrast, the authors of [392] considered a scenario of ISAC-UAV with multiple users and eavesdroppers. They optimized transmit BF and the UAV positioning to minimize the maximum eavesdropper's SINR under the constraints of transmit power, communication user's QoS, and sensing BPG. The authors of [391] considered a similar scenario with a jamming UAV. Based on the target location information estimated by the ISAC-UAV, the jamming UAV disseminates jamming signals to the eavesdroppers. A joint optimization of user scheduling, transmit power, and UAV trajectory was conducted to achieve the maximal SR.

In [393], [401], the ISAC-UAV model has been examined in the context of RIS-aided networks, considering the scenario with multiple targets and a separate eavesdropper whose CSI is not accessible. The authors have proposed a joint optimization framework, encompassing the transmit power allocation, the scheduling of users and targets, the phase shifts of the RIS, and the trajectory and velocity of the UAV. This framework aimed to maximize both the average achievable rate and the energy efficiency.

Additionally, the authors of [394] proposed a high-altitude platform station (HAPS)-mounted FD ISAC system, where a ISAC BS simultaneously serves communication users, senses multiple targets, and offloads part of the sensed data to a dedicated edge server. The deployment of an UAV is also considered, with the primary function of transmitting jamming signals to disrupt the communication links with malicious targets or potential eavesdroppers identified through the sensing process. The authors formulated a multi-objective optimization problem is formulated to jointly optimize the transmit/receive BF and UAV trajectory to maximize the communication rates of users under the constraints of radar SINR, the maximum SINR of the eavesdropper, offloading, and power constraints.

*Summary:* The utilization of UAVs in three-dimensional space, along with their robust air-ground LOS channels, is expected to enhance communication and sensing coverage, optimize surveillance flexibility, and improve communication and sensing performance in comparison to terrestrial ISAC systems. Nevertheless, this novel aerial ISAC paradigm concomitantly introduces new design challenges [402]. Specifically, the increased vulnerability to eavesdropping and jamming attacks due to the LOS-dominated airground channels poses a significant challenge to the effective safeguarding of legitimate communication and sensing users in UAV-ISAC. Furthermore, while the existing literature focuses on a single UAV, the use of multiple UAVs to collaboratively provide ISAC services has been identified as an efficient solution to further enhance the communication and sensing coverage. However, such systems necessitate more sophisticated interference management [403].

## J. Covert Communication

In contrast to conventional secure transmission methods, covert communication involves the transmission of messages

TABLE XV
SUMMARY OF PAPERS ON COVERT COMMUNICATIONS FOR ISAC.

| Reference | User(s) | Eve(s) | Target(s) | KPIs (Com. / Sen.) | Eve's CSI | Opt. Variables | Optimization Techniques |
|---|---|---|---|---|---|---|---|
| [404] | S | S | S | CR / MI | Perfect / Imperfect | BF | Bisection Search, SDR, S-Procedure, ZF |
| [405] | S | S | S | CR / MI | Perfect | BF, Power Allocation | SDR |
| [406] | S | S | S | CR / MI | Perfect | BF, RIS Phase Shifts | AO, SDR |
| [407] | S | S | S | CR / MI | Imperfect | BF, STAR-RIS Phase Shifts | AO, Penalty CCP, SCA, Schur Complement, Sign-Definiteness, S-procedure |
| [408] | S | S | S | CR / TDP | Perfect / Imperfect | BF | SDR, S-Procedure |
| [409] | S | M | M (Warden) | CR / BPM and CC | Imperfect / Statistical | BF, AN | BTI, DC Relaxation, S-Procedure |
| [410] | S | S | S (Warden) | CR / Tracking MSE | Perfect / Imperfect | BF, Radar Waveform | FP, Schur Complement, SDR |
| [411] | S | S | S (Warden) | CR / TDP | Unknown | Transmit Waveform, AN Power, Transmission Power / Rate, Detection / Communication Trade-off | AO |
| [412] | M | S | S | CR / SINR | Perfect | BF, Radar Receive Filter | AO, FP, Penalty Method, SDR |
| [413] | M | S | S | CR / SINR | Perfect | BF | Bisection Search, FP, SDR |
| [414] | M | S | S | CR / SINR | Statistical | BF, Radar Waveform, Radar Receive Filter, STAR-RIS Phase Shifts | AO, FOTE, SCA, SDR, SROCR |
| [415] | M | S | S | CR / SNR | Perfect | Power Allocation, UAV Trajectory Communication Scheduling | AO, FOTE, SCA |
| [416] | M | S | M | CR / CRB | Statistical | BF, RIS Phase Shifts | AO, FP, Penalty Method |
| [417] | M | M | M (Warden) | CR / BPM | Imperfect | BF | Penalty Method, SDR, SCA |
| [418] | M | M | M (Warden) | CR / BPM | Imperfect | BF, Radar Waveform, Radar Receive Fitler | AO, BTI, S-Procedure |

S: Single, M: Multiple.

through wireless networks without being detected by the intended recipient (warden). This is achieved by concealing the transmission within environmental or artificial noise, thereby providing a higher level of security and privacy compared to cryptography and other PLS techniques [77], [419], [420]. The field of covert communication has been extensively studied in a wide range of wireless communication applications, including UAVs [421]–[424], RISs [425]–[430], and ISAC systems. A summary of papers that consider covert communications for ISAC is provided in Table XV, and a review of these papers is presented below.

*1) Single-User Scenario:* The paper [404] is among the early works that explored covert communication in ISAC systems, proposing a transmit BF design that aims to optimize both the CR and the radar MI. Building upon this foundation, they extended the BF design to the context of imperfect CSI of the warden, leading to the proposal of a robust BF design. Similarly, the communication and sensing performance metrics, such as CR and the radar MI, are also considered in a jammer-assisted ISAC system [405], passive and active RIS-aided ISAC systems [406], [407], respectively. In contrast to the radar MI, the authors in [408] considered the TDP as a sensing metric. In addition to imperfect CSI, the authors in [409] considered statistical CSI of the warden and proposed robust BF to minimize the sensing metric, i.e., a weighted sum of BPG MSE and CC, under the constraints of a CR and the transmit power budget.

Meanwhile, the authors of [410] proposed a sensing-aided covert communication system for a radar-communication co-operation system in the presence of an aerial adversary target, which acts as a warden. They employed the EKF to track and predict the trajectory and the corresponding channels of the adversary target based on the radar echoes from the target. They proposed a joint design of transmit BF and radar waveform to maximize the CR under the tracking MSE constraint, considering both perfect and imperfect CSI of the adversary target.

More recently, drawing the inspiration from the work in [31], [431], the authors of [411] have proposed a two-stage framework with the objective of fully leveraging the sensing function of ISAC in the context of covert communication. This framework is illustrated in Fig. 12. Specifically, in the initial stage (Fig. 12a), a FD ISAC transceiver, designated as Bob, performs sensing to detect the warden while transmitting uplink signals (e.g., about the warden's presence) to another transceiver, Alice. In the subsequent stage (Fig. 12b), Alice covertly transmits downlink signals to Bob while Bob generates AN based on his detection outcome in the initial stage. Their simulation results demonstrated that the sensing function of ISAC offers a substantial covert communication performance enhancement.

*2) Multi-User Scenario:* The authors of [412]–[416] considered a scenario involving a single warden. Specifically, in [412], the authors considered a secure IoT system in which a BS disseminates a probing waveform to detect an IoT sensing target in the presence of clutters while simultaneously serving public IoT communication users and sending confidential communication data to a covert IoT user. In this

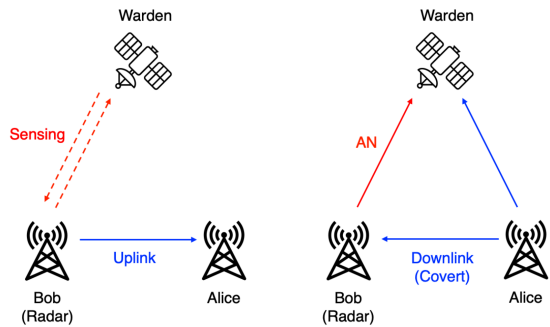

(a) First step: Bob performs sensing while transmitting information about warden's presence to Alice.

(b) Next step: Alice covertly transmits signals to Bob while Bob generates AN to interfere with warden's detection.

Fig. 12. Two-stage method for sensing-assisted covert communication in [411].

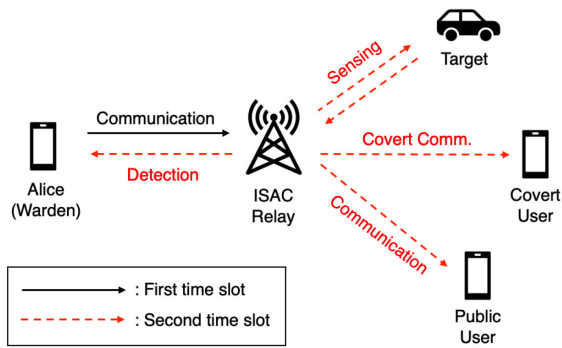

Fig. 13. A scenario of covert communication in cooperative NOMA-aided ISAC in [413].

scenario, they optimized the transmit BF and the radar receive filter, with the objective of maximizing the CR, subject to constraints of the radar SINR and the communication QoS of public users. They developed algorithms for designing two different BF architectures, i.e., fully digital BF and hybrid BF. Additionally, the authors of [413] considered a cooperative NOMA system illustrated in Fig. 13. In this system, Alice transmits signals to the ISAC relay in the first time slot, and subsequently, the ISAC relay attempts to transmit covert signals to a covert user without being detected by Alice while simultaneously forwarding Alice's signals to a public user and sensing the target. In this configuration, they derived a minimum detection error probability for the warden under the assumption that the detection threshold can be dynamically adjusted by the warden. They jointly optimized covert/public communication BF and sensing BF to maximize the CR under the NOMA decoding order requirement, the QoS requirement of the public user, the sensing/covert constraints, and the total power budget. Furthermore, the authors of [415] considered a UAV-ISAC system in the presence of a single target and warden. They proposed a joint optimization of the power allocation for communication and radar signals, as well as trajectory and communication scheduling of the UAV-ISAC, with the objective of maximizing the covert rate.

In [416], the authors considered an active RIS-assisted NOMA-ISAC system, wherein the ISAC BS serves two NOMA users while sensing multiple mobile targets. To optimize the CR, the transmit BF and the active RIS reflection are jointly optimized, contingent upon the QoS requirements of the NOMA public user, the constraint of CRB, and the covertness level of the warden. In addition, the authors of [414] considered the use of an active STAR-RIS [432] for an ISAC system in the presence of both an eavesdropper and a warden. The ISAC BS serves both communication and covert users while sensing a target, which acts as an eavesdropper. To maximize the average sum of the minimum secrecy and CRs while satisfying the constraints of the sensing SINR and the transmit power at the BS and the active STAR-RIS, they jointly optimized transmit BF and receive filter at the BS and the transmission/reflection BF at the active STAR-RIS.

In contrast to the aforementioned works, which consider a single warden, multiple wardens are considered in the papers [417], [418]. The authors of [418] proposed robust BF for a covert ISAC system under both bounded and probabilistic CSI models. They formulated the worst-case and outage-constraint robust design of joint transceiver BF and radar waveform. Conversely, the authors of [417] examined the scenario of multiple colluding and non-colluding wardens under sensing-assisted imperfect CSI of wardens. They proposed a robust BF design aimed at maximizing the CR, subject to the constraint of MSE between the desired and designed beams.

*Summary:* The preceding studies have demonstrated that the sensing functionality of ISAC enhances covert communication performance. Furthermore, certain studies considered practical scenarios with imperfect CSI of the warden(s) as well as legitimate receivers. In addition, it has been demonstrated that the implementation of RIS and NOMA is effective in achieving higher CRs. Conversely, one of the prospective future directions is the consideration of active wardens. The extant literature typically assumes a fixed warden or one that engages in passive detection. In contrast, an active warden may modify its detection location and the detection strategy in response to the received signals [77]. While the authors of [410] considered a single mobile target (warden) and a single communication user, extensions to more complex scenarios with multiple mobile targets and communication users are of practical importance.

### K. Semantic Communications

Recent advancements in deep learning technologies have prompted the development of deep learning-enabled end-to-end semantic communication systems, which have garnered significant attention [437]–[446]. In contrast to conventional communications, which prioritize the accurate and effective transmission of bits, semantic communications employ semantic data extraction to convey information, emphasizing the interpretation and understanding of data.

The integration of ISAC with semantic communication represents a novel research subject with a paucity of related publications [447]. Specifically, we have summarized papers

TABLE XVI
SUMMARY OF PAPERS ON SEMANTIC COMMUNICATIONS FOR ISAC.

| Reference | User(s) | Eve(s) | Target(s) | KPIs (Com. / Sen.) | Eve's CSI | Opt. Variables | Optimization Techniques |
|-----------|---------|--------|-----------|--------------------|-----------|----------------|--------------------------|
| [433] | M | S | S | SSR / CRB | Perfect | BF, AN | Golden-Section Search, SDP, Schur Complement |
| [434] | M | M | M (Eve) | SSR / CRB | Imperfect | BF, Semantic Extraction Ratio | AO, FOTE, SDR |
| [435] | M | M | M (Eve) | SSR / BPM | Perfect | BF, Semantic Extraction Ratio | AO, FOTE, SDR |
| [436] | M | M | M | SSE / CC | Perfect | BF | AO, FOTE, FP |

S: Single, M: Multiple, SSE: Secure Semantic Efficiency.

that address PLS problems in semantic communication for ISAC systems in Table XVI.

In [433], the authors considered an integrated sensing and semantic communication system wherein ISAC BS serves multiple semantic communication users while monitoring a malicious sensing target in the presence of an eavesdropper. They employed AN and a dedicated sensing signal to augment PLS and derived the CRB and a semantic secrecy rate (SSR) based on the semantic rate in [448] as KPIs for sensing and communication, respectively. To achieve a trade-off between these KPIs, they solved a multi-objective optimization problem using SDP, Gaussian randomization, and golden section methods. In contrast, the authors of [436] considered multiple eavesdroppers and targets. They adopted the same semantic similarity-based definition of the semantic rate as [433]. Then, they attempted to maximize the secure semantic efficiency, which is defined as the ratio of the sum SSR to the total transmit power, by jointly designing transmit BF and a semantic communication parameter.

In [434], [435], [449], the authors introduced an alternative definition of SSR defined as the number of bits received by users after de-extracting semantic information. In [435], [449], they considered a joint design of transmit BF and a semantic extraction ratio to maximize the sum SSR of all users while maintaining the minimum QoS for each user and guaranteeing overall sensing performance. Furthermore, a weighted optimization is formulated to achieve trade-offs between the sum SSR and CRB in [434].

*Summary:* While several papers have considered a SSR, the appropriate definition of the KPI for PLS in semantic communication remains an open question. In [433], [436], the authors adopted the semantic rate based on the semantic similarity [448] as a performance measure of text semantic communications. However, it should be noted that no closed-form expression for the semantic similarity, which characterizes the similarity between an original data set and the corresponding recovered data set, currently exists. To address this issue, a data regression method was proposed in [450] for text semantic communications, which approximates the semantic similarity by a generalized logistic function. Nevertheless, the development of a unified performance metric for multimodal semantic communication (e.g., image, video, and speech) remains an important future work.

*L. Others*

In the following, we conduct an examination of existing literature related to PLS for ISAC in other scenarios. A summary of these papers is outlined in Table XVII.

CSI-based sensing is a common method for monitoring human activities [459] and a malicious nose could exploit the estimated CSI to sense the surrounding environment and extract information about a specific target [460]. To address this issue, the authors of [451] proposed an adaptive pilot allocation scheme for OFDM ISAC systems by exploiting the reciprocal channel between legitimate users in time division duplex (TDD) systems to hide the pilot location information from an eavesdropper. While the sensing functionality of ISAC was not a primary focus in this paper, the security performances were evaluated in terms of pilot location error rate, MSE of estimated channels, and Eve's BER degradation with respect to Bob's BER.

In [452], the authors considered a frequency hopping (FH) MIMO radar-based DFRC system and revealed the potential of employing permutations of hopping frequencies to enhance security. Specifically, in addition to hopping frequency combination selection (HFCS) [461], they performed hopping frequency permutation selection (HFPS) and developed the element-wise phase compensation method to demodulate HFPS signals. It was demonstrated that, given the requirement of the user's angle-of-departure (AoD) information for the demodulation of HFPS signals, the proposed scheme significantly degrades the eavesdropper's SER relative to the user's SER.

In [458], the authors explored a joint design of the transmission and radar sensing power to enable PLS in the context of ISAC-based internet of vehicles (IoV) systems. Specifically, they formulated a SR maximization problem, subject to constraints related to communication outage probability and sensing accuracy in terms of success ranging probability. Success ranging probability was defined as the probability that the received SINR is greater than or equal to a certain threshold.

In [457], the authors explored the potential of high-speed train-to-ground communication, leveraging the sensing capability of ISAC to detect eavesdroppers and the high directivity of millimeter wave (mmWave) to generate signal shadow regions at the eavesdropper's location. This approach ensures an exceptionally low signal reception power for the eavesdropper. They jointly optimized transmit waveform and BF to maximize the sum-rate of the mobile relays on the train roof while

TABLE XVII
Summary of papers on other related scenarios.

| Reference | User(s) | Eve(s) | Target(s) | KPIs (Com. / Sen.) | Eve's CSI | Opt. Variables | Optimization Techniques |
|-----------|---------|--------|-----------|--------------------|-----------|----------------|--------------------------|
| [451] | S | S | - | BER / - | Unknown | - | - |
| [452] | S | S | S | SER / Ambiguity Function [453] | Unknown | - | - |
| [454] | S | S | S | MI / Distortion | Unknown | - | - |
| [455] | S | S | S | MI / Distortion | Unknown | - | - |
| [456] | S | S | S | MI / Distortion | Unknown | - | - |
| [457] | M | M | M | User's Rate / CRB | Unknown | BF, Transmit Waveform | AO, SCA |
| [458] | S | S | S | SR / SRP | Unknown | Transmission / Sensing Power | AO, FOTE, FP, SCA |

S: Single, M: Multiple, SRP: Success Ranging Probability.

ensuring the sensing performance in terms of CRB and the creation of shadow regions based on the concept of *absolute security* [462]. They developed an AO algorithm based on the SCA technique and demonstrated through simulations the efficacy of the combination of ISAC and absolute security.

Meanwhile, the authors of [454] examined the secure ISAC model from the information-theoretic perspective. Specifically, they derived inner and outer bounds on the secrecy-distortion region for a discrete memoryless state-dependent broadcast channel model. This model introduces an eavesdropper to the model in [463], in which messages are encoded and sent through a state-dependent channel with generalized feedback both to reliably communicate with a receiver and to estimate the channel state by using the feedback and transmitted codewords. The work [454] has been extended to the action-dependent setup by introducing transmitter actions that affect the channel states in [455]. Furthermore, the authors of [456] have established an achievable secrecy-distortion region for stochastically degraded secure ISAC channels under bivariate Rayleigh fading.

*Summary:* A considerable body of research has centered on the design, analysis, and optimization of pragmatic ISAC technologies for diverse ISAC systems, including OFDM ISAC and FH ISAC. Concurrently, it is imperative to examine the fundamental limitations of ISAC to elucidate the discrepancy between the state-of-the-art technologies and their performance boundaries. This examination will furnish valuable insights and guidance for the advancement of enhanced ISAC technologies that can approach these performance limits [6], [9]. In this regard, it is important to investigate the fundamental limits of ISAC under emerging scenarios, such as those involving RIS, along with the practical considerations, including channel estimation error and frequency offset.

## IV. Future Research Directions

In the following, we explore prospective avenues for research in the related topic of secure ISAC.

### A. Sensing Security

PLS techniques in ISAC for ensuring communication security have been thoroughly examined in the existing literature, which is the primary focus of the present paper. However, it is imperative to acknowledge that ISAC systems are also susceptible to novel sensing security threats. This is due to the fact that the sensing information may be vulnerable to sensing eavesdroppers. The aspect of sensing security in ISAC has not been extensively investigated in the literature, with the exception of some recent papers, e.g., [17], [241], [464]–[466].

The ability of ISAC systems to leverage sensing signals enables the covert interception of sensing results without the need for active transmission of their own signals. This capability allows adversaries to infer the actions of associated physical systems and potentially launch further actions that could compromise system performance. Notably, this passive eavesdropping on sensing information lacks protection from higher-layer security approaches, introducing additional privacy and security challenges to the design of physical layer tailored for ISAC. Consequently, ISAC BF design, in conjunction with other key technologies, such as RIS and UAV, must be formulated to thwart both data eavesdropping and adversary sensing.

### B. Practical Communication-Centric Design

In comparison with sensing-centric and joint designs of ISAC, the communication-centric design, which incorporates sensing functionality into existing communication waveforms, holds greater promise for practical implementation in next-generation wireless networks due to its compatibility with existing infrastructure [467]. Recently, the achievable sensing performance of communication-centric ISAC signals employing random signaling has been investigated by designing communication systems [117], [468]–[470]. In [467], [471], design guidelines for core building components in communication-centric ISAC were provided, including modulation, constellation, pulse shaping, and MIMO precoding, for minimizing the average squared auto-correlation function of random ISAC signals.

However, the majority of existing literature on PLS for ISAC operates under the assumption of ideal capacity-achieving coding and i.i.d. Gaussian signaling. The impact of practical coding and modulation, as well as pulse shaping, on the resulting performance has been disregarded. Consequently, the system-level design of communication-centric ISAC for PLS holds practical significance. This encompasses the design of channel coding (e.g., low-density parity check (LDPC) and polar codes) [472], [473] and constellation (e.g., adaptive coding modulation and constellation shaping) [474]–[476], in conjunction with symbol-level precoding in Section III-F for achieving a trade-off between sensing and PLS performance.

## C. Network-Level ISAC

Network-level ISAC refers to the collaboration of multiple ISAC transceivers across different cells to enhance both wireless communication and sensing [477]–[480]. The ISAC network, such as CF networks in Section III-B, can cover larger surveillance areas than single-cell ISAC, since each ISAC BS may also receive the target-reflected signals transmitted by multiple BSs or users, as multi-static sensing [481]. Furthermore, joint BS selection and BF design in ISAC networks is a promising approach to significantly improving the information/sensing security. Therefore, it is critical to design a cooperation framework for striking a trade-off between the sensing and communication security performances [477].

Conversely, the network ISAC confronts novel technical obstacles concerning wireless resource allocation and user/target scheduling. Specifically, interference resulting from concurrent transmissions by distributed BSs has the potential to substantially compromise the sensing and PLS performance within the network. To address this challenge, it is imperative to undertake a thorough examination of cooperative strategies, including ISAC network synchronization, coordinated BF, and cooperative resource allocation.

## D. AI-Aided Efficient Optimization

The existing literature has demonstrated that the integration of ISAC with other key technologies, such as RIS and UAV, will enhance its PLS and sensing performance by providing additional optimization DoF. However, such systems may suffer from the prohibitively high computational complexity associated with the system design and optimization, and thus conventional time-consuming iterative optimization algorithms may not be applicable. Specifically, in the context of a RIS-assisted network, the discrete phase shifts of the RIS and the consideration of practical constraints, such as hardware and channel estimation imperfections, give rise to complicated non-convex optimization problems. Furthermore, for scenarios involving multiple RISs and wideband signals, the dimension of the variables to be optimized will be substantial [285]. A similar challenge arises in UAV-enabled ISAC networks, where the problem of dynamic deployment and resource allocation of UAVs is a crucial concern [402].

To address these challenges, AI is anticipated to deliver more efficient and robust solutions [482]. For instance, deep learning can be employed to offer low-complexity solutions to the joint design of ISAC waveform and RIS reflection/UAV trajectory. Specifically, DRL is well-suited for handling the control of the RIS/UAV in dynamic heterogeneous networks by integrating sensing information, such as the states of the environment, into the algorithm [483].

## E. Movable Antenna-Based ISAC

MA has emerged as a promising technology to enhance the performance of wireless communication by enabling antenna movement [291], [484]–[486][7]. Unlike conventional FPAs undergoing random fading channels, MAs can dynamically change their placements to improve channel conditions, often with a reduced number of antennas and RF chains. The transition from conventional FPA to MAs represents a paradigm shift with the potential to unlock significant opportunities for the evolution of next-generation wireless networks, such as 6G [488], [489].

The application of MAs in ISAC systems has recently undergone extensive study [490]–[497]. It has been demonstrated that, in comparison with conventional FPA arrays, the new DoF of MA arrays via antenna position optimization has the potential to significantly improve communication and sensing performance. However, despite the potential advantages, the extension of MAs to PLS problems remains in its infancy. A notable challenge in the design of MA-assisted ISAC systems pertains to the joint optimization of MA element positions and BF, which typically leads to a non-convex optimization problem. A promising solution to address this intractable optimization problem involves the integration of deep learning techniques, as demonstrated in [492]. Another significant challenge in MA-enabled ISAC systems arises from the difficulty of channel estimation, attributable to the continuous dynamic changes in the channel caused by the movement of MA elements.

## F. Emerging Metasurface Technologies for ISAC

In Section III-D, we explored the potential of RIS deployment in propagation environments to enhance secure communication and sensing. Meanwhile, reconfigurable metasurfaces have emerged as a promising solution for active transceiver applications, enabling the concept of holographic MIMO (HMIMO) communications [498]–[500]. These metamaterial-based antennas possess the capability to modulate EM waves through the use of simple diode-based controllers, thereby demonstrating their potential to offer both energy-efficient and cost-effective solutions. The two most representative enablers of HMIMO are dynamic metasurface antennas (DMAs) [501] and reconfigurable holographic surfaces (RHSs) [502], [503].

In addition, recent proposals have introduced three-dimensional (3D) stacked intelligent metasurfaces (SIMs) as a novel approach to analog signal processing in the wave domain. These SIMs are constructed by stacking multiple programmable metasurface layers, which bear conceptual similarities to an artificial neural network [504], [505]. By meticulously configuring the transmission coefficients of the meta-atoms on each metasurface, SIMs can autonomously perform various computations and signal processing tasks without requiring digital storage, transmission, or external computing power. Furthermore, a flexible intelligent metasurface (FIM) has emerged as a soft array consisting of several low-cost radiating elements, each of which can independently emit EM signals. These radiating elements can flexibly adjust their position, even perpendicularly relative to the overall surface, thereby enabling the metasurface to morph its 3D shape [506], [507]. While some studies have explored the application of these metasurface technologies in the context of ISAC [508]–[510], investigating their potential for PLS problems within the broader framework of ISAC represents a promising avenue for future research.

---

[7]Despite their different origins and implementation, MAs are conceptually similar to fluid antenna systems (FASs) in exploiting flexible antenna positioning and can be considered interchangeable terms [487].

## V. Conclusion

The ISAC system is envisioned to enhance the performance and capabilities of 6G networks by integrating wireless sensing and communication functions in an efficient manner. However, it is important to acknowledge that ISAC introduces distinctive security vulnerabilities within the wireless network. This paper has provided a comprehensive survey of recent advances in PLS techniques in ISAC systems, with the aim of addressing these unique security vulnerabilities. Initially, we conducted a thorough examination of several key performance metrics for sensing and PLS, along with a range of optimization techniques that are extensively adopted to address PLS design problems for ISAC systems. Subsequently, we conducted a comprehensive review of the existing PLS schemes for ISAC, categorizing them based on the considered scenario to highlight the current research landscape. Finally, we discussed several research directions in PLS techniques for ISAC systems.